\documentclass[prd,aps,nofootinbib,notitlepage,showpacs,showkeys,preprintnumbers]{revtex4}
\usepackage{graphicx,epsf,amsmath,amsfonts,amssymb,amsbsy}
\usepackage{epsfig}
\usepackage{epstopdf}

\newcommand{\ds}{\displaystyle}
\newcommand{\vev}[1]{\langle#1\rangle}
\newcommand{\mat}{\left ( \begin{array}}
\newcommand{\emat}{\end{array} \right )}
\newcommand{\vect}{\left ( \begin{array}{c}}
\newcommand{\evect}{\end{array} \right )}

\newcommand{\Det}{\mathop{\rm Det}\nolimits}

\begin{document}

\title{ \bf Electrical neutrality and $\beta$-equilibrium conditions in dense quark matter: generation of charged pion condensation by chiral imbalance
}
\author{T. G. Khunjua $^{1),~2)}$, K. G. Klimenko $^{3)}$, and R. N. Zhokhov $^{3),~4)}$ }

\affiliation{$^{1)}$ The University of Georgia, GE-0171 Tbilisi, Georgia}
\affiliation{$^{2)}$ Department of Theoretical Physics, A. Razmadze Mathematical Institute, I. Javakhishvili Tbilisi State University, GE-0177 Tbilisi, Georgia}
\affiliation{$^{3)}$ State Research Center
of Russian Federation -- Institute for High Energy Physics,
NRC "Kurchatov Institute", 142281 Protvino, Moscow Region, Russia}
\affiliation{$^{4)}$  Pushkov Institute of Terrestrial Magnetism, Ionosphere and Radiowave Propagation (IZMIRAN),
108840 Troitsk, Moscow, Russia}

\begin{abstract}

The phase diagram of dense quark matter with chiral imbalance is considered with the conditions of electric neutrality and $\beta$-equilibrium. It has been shown recently that chiral imbalance can generate charged pion condensation in dense quark matter
, so it was interesting to verify that this phenomenon takes place in realistic physical scenarios such as electrically neutral matter in $\beta$-equilibrium, because a window of pion condensation at dense quark matter phase diagram (without chiral imbalance) predicted earlier was closed by the consideration of these conditions at the physical current quark mass. In this paper it has been shown that the charged pion condensation is generated by chiral imbalance in the dense electric neutral quark/baryonic matter in $\beta$-equilibrium, i. e. matter in neutron stars. It has been also demonstrated that pion condensation is inevitable phenomenon in dense quark matter with chiral imbalance if there is non-zero chiral imbalance in two forms, chiral and chiral isospin one. 
It seems that in this case pion condensation phase can be hardly avoided by any physical constraint on isopin imbalance and that this conclusion can be probably generalized from neutron star matter to the matter produced in heavy ion collisions or in neutron star mergers. The chiral limit and the physical piont (physical pion mass) has been considered and it was shown that the appearance of pion condensation is not much affected by the consideration of non-zero current quark mass.
\end{abstract}


\maketitle

\section{Introduction}
It is expected that there exists a rich phase structure of QCD at finite temperature and baryon density. For example, lattice simulations from the first principles have revealed that the confined quarks will become released to quark-gluon plasma around the temperature $T_{c}=\Lambda_{QCD}$. The study of hot and dense QCD is not just pure academic research but is important for various physical applications ranging from the evolution of the early universe through neutron star physics to heavy-ion collisions. It is of interest how quarks and gluons behave if the system is heated up or is compressed. In heavy-ion collision there exist a fireball of very hot matter as a rule with small baryon chemical potential $\mu_B$. The region of phase diagram with high  $\mu_B$ and almost zero temperatures could be realized in nuclear/quark matter in the interior of compact stars.

In the last years lattice QCD as an ab initio approach has made huge progress in describing the thermodynamics of QCD and has provided important contributions to the understanding of the QCD phase diagram.
However, computer power is not the only limitation of the lattice approach. While it is the feasible method in the case of zero baryon chemical potential, it has a conceptual problem when calculating at $\mu_B\ne 0$. For finite chemical potential $\mu_B$ the fermion determinant, which is used as
a probability weight in the Monte Carlo sampling, becomes complex and thus the method fails. This is known as the sign problem. Although there are ways to explore the region of small chemical potential $\mu_B$ on lattice, but up to now lattice calculations at intermediate (let alone the large) values of $\mu_B$ is not possible.

Therefore, effective QCD-like models which exhibit the features of color confinement and spontaneous chiral symmetry breaking are more feasible to be used to study the phase structure of QCD with nonzero density in addition to temperature. One of the most widely used effective model is the Nambu--Jona-Lasinio (NJL) model \cite{Nambu:1961fr} (see for review \cite{Klevansky:1992qe,Hatsuda:1994pi,Buballa:2003qv}). The degrees of freedom of this model are not hadrons, as in the chiral perturbative theories, but self-interacting  quarks  and there are no gluons in considerations, they are integrated out. The model is tractable and can be used as low energy effective theory for QCD.  The most attractive feature of the NJL models is the dynamical breaking of the chiral symmetry  (quarks acquirement of a comparatively large mass)  and hence it can be used as a basis model for constituent quark model.

There are additional parameters besides the temperature and the baryon density (or baryon chemical potential $\mu_B$) which may be relevant for the above mentioned dense QCD matter. They
are, for instance, isospin $\mu_I$ and chiral $\mu_{5}$ chemical potentials. The former is used to describe a medium with isospin imbalance, when there are unequal densities of $u$ and $d$ quarks. While the second is introduced in order to take into account the chiral imbalance, i.e. different average numbers of right-handed and left-handed quarks. 
There are two types of chiral imbalance that are described by chiral chemical potential $\mu_{5}$ and chiral isospin chemical potential $\mu_{I5}$.  There are a lot of works \cite{andrianov} that considered chiral imbalance $\mu_{5}$  in quark matter and it is thought that it can be generated in the fireball after heavy-ion collision as a consequence of the famous chiral Adler-Bell-Jackiw anomaly, and it plays an important role in the chiral magnetic effect phenomenon \cite{Fukushima:2008xe}. 
Moreover, chiral densities 
$n_5$ and $n_{I5}$ (and hence $\mu_{5}\ne 0$ and $\mu_{I5}\ne 0$) can be produced in dense quark matter in the presence of external magnetic field \cite{kkz3, kkz2}  due to the so-called chiral separation effect \cite{Metlitski} or in fast rotating dense matter due to the so-called chiral 
vortical effect \cite{Fukushima:2018grm}. Chiral imbalance $n_5$ can also be produced in parallel 
electric and magnetic fields \cite{Ruggieri:2016fny}. Due to the different quark electric charges, equally well in parallel 
electric and magnetic fields can be produced $n_{I5}$ \cite{kkz2}. 
The influence of the chiral imbalance $\mu_{I5}$ on the phase structure has been considered in \cite{kkz, kkz2, kkz3, ekk, Khunjua:2019nnv, Hongo:2018cle, Sogabe:2019gif, Chao:2018ejd, Chao:2018czo, Thies:2019ejd}. Let us also note that the inclusion of chiral isospin chemical potential $\mu_{I5}$ is formally more rigorous \cite{kkz2, kkz3} and does not have the problems of $\mu_{5}$.

In connection with the physics of neutron star matter and experiments on heavy-ion collision, there has recently appeared an interest in the study of quark medium with isospin (isotopic) asymmetry. QCD phase diagram at nonzero values of the isotopic chemical potential $\mu_I$ has been studied in different approaches, e.g., 
in lattice QCD approach \cite{Kogut, Brandt:2017zck}, in different QCD-like effective models such as NJL model \cite{koguthe,eklim,eklim1, abuki,ak,aryal,Khunjua:2019nnv, Chao:2018ejd, Cao:2015cka}, quark meson model \cite{Folkestad:2018psc}, ChPT \cite{Adhikari:2019mdk, Son}, also in perturbative QCD (with
diagrams resummation) \cite{Andersen:2015eoa}, in effective theory at
asymptotically high isospin density \cite{Cohen:2015soa}, in a random matrix model \cite{Klein:2003fy}, in hadron resonance gas model \cite{Toublan:2004ks}.
It was shown in these papers that if there is an isospin imbalance then charged pion condensation (PC) phenomenon can be generated in quark matter. Recent review on meson condensation including pion condensation is \cite{Mannarelli:2019hgn}. Even idea about pion stars, type of compact star whose main constituent is a Bose-Einstein condensate of charged pions, has been discussed \cite{Andersen:2018nzq, Brandt:2018bwq}

 The possibility of pion condensation in a nuclear medium was speculated a long time ago, in the early 1970s, by Migdal and others \cite{Migdal, Sawyer:1972cq}. 
 In \cite{Migdal} the neutral pion condensation has been considered, the charged  pion condensation in neutron rich matter has been investigated for the first time in \cite{Sawyer:1972cq, Baym:1975mf}. 
 After that many efforts have been made to clarify these phenomena as a better insight into such condensation phenomena would lead to important advances 
 in 
pionic atoms physics
\cite{Toki:1989wq, Kienle:2004hq}, in the physics of neutron stars \cite{Hartle:1974de, Maxwell:1977zz}, supernovae \cite{Ishizuka:2008gr}, and heavy ion collisions \cite{Zimanyi:1979ga}.
In nuclear matter crucial role in the study of pion condensations is played by the Landau-Migdal (LM) parameters for nucleon-nucleon, nucleon-$\Delta$ and $\Delta$-$\Delta$ couplings.
In most of the earlier calculations universality was assumed (LM parameters are equal) and a rather large
value of LM parameter was obtained, for which pion condensations are hardly expected even at high densities \cite{Muto:2003dk, Suzuki:1999hn}.
Then new information on the LM parameters has been obtained from the
experiments on the giant Gamow-Teller (GT) resonances and the results show that the critical density of 
neutral PC in symmetric
nuclear matter and that of 
charged PC in neutron matter have been estimated
to be of the order or less than two normal nuclear density for both condensations  \cite{Muto:2003dk,Suzuki:1999hn, Ichimura:2006mq}. 
Similar results has been obtained in the new experiments on GT resonances \cite{Yasuda:2018den}.
The situation is still not clear and further studies on more neutron-rich nuclei are needed and planned \cite{Yasuda:2018den}.

All said above concerned p-wave pion condensation, possibility of s-wave pion condensation (zero momentum condensation)  in neutron stars has been explored in \cite{Ohnishi:2008ng}.
 It has been shown that s-wave pion condensation is very unlikely to take place in neutron stars. Proton fraction and electron chemical potential in \cite{Ohnishi:2008ng} was evaluated in the hadronic relativistic mean field models and, in general, all the studies discussed above were performed in hadronic models and was mostly concerned with pion-nucleon interaction,  coupling between pions and baryons in the nuclear medium and considered the pion as an elementary object. 
 Then this gap was filled and the possibility of s-wave charged pion condensation was studied in microscopic model built of quarks and a composite pions (with quarks as the constituents), namely in NJL model, which exhibits chiral symmetry restoration at the finite value of quark chemical potential or temperature \cite{koguthe, eklim, eklim1}. It was also shown that there could be PC in dense baryonic matter with non-zero baryon density \cite{eklim} and it was even shown in \cite{eklim1} that this phase is realized in electric neutral dense matter that is in $\beta$-equilibrium. This was in a slight 
 Then it was shown that charged pion condensation found in \cite{abuki} is extremely fragile to explicit chiral symmetry breaking by a finite current quark mass and in the physical point (with the physical current quark mass) it completely evaporates. So one can conclude that as in the framework of hadronic models as in the framework of quark models  in a pure electric neutral dense hadron/quark 
 matter in $\beta$-equilibrium (without additional conditions such as, for example, chiral imbalance) there is no pion condensation phenomenon.

However, several external factors have recently been discovered which promote the formation of the charged PC phase in dense quark matter \cite{Khunjua:2019nnv}. \footnote{Among these factors, the finite size of the system \cite{ekkz}, the spatial inhomogeneity of its pion condensate \cite{gkkz} as well as its chiral imbalance \cite{ekk, kkz}.} One of them is chiral imbalance, namely the chiral $\mu_{5}$ and the chiral isotopic $\mu_{I5}$ (in other words, an alternative form of the chiral imbalance of a system) chemical potential $\mu_{I5}$. The effect of chiral isospin imbalance on dense quark medium was considered in the chiral limit both within the framework of a toy NJL$_2$ (in Ref. \cite{ekk}) and in a more realistic NJL$_4$ (in Refs. \cite{kkz,kkz2,kkz3}) models. The effect of chiral  chemical potential has been considered in \cite{kkz}. It has been shown that chiral imbalance leads to generation of pion condensation in dense quark matter. But in these papers the influence of chiral imbalance on the formation of the charged PC phenomenon  was considered without taking into account the possible electric neutrality and $\beta$-equilibrium of dense quark matter and the real nonzero values of $m_0$, i.e. the results of these studies are still not applicable to such astrophysical objects as neutron stars, etc.

In the present paper we fill this gap of our previous researches of charged PC phase of dense quark medium and study the phase diagram of QCD and charged PC phenomenon in the framework of (3+1)-dimensional  NJL model with two quark flavors and in the presence of the 
chiral isospin $\mu_{I5}$, chiral $\mu_5$, etc chemical potentials under electric charge neutrality and $\beta$-equilibrium constraints and with physical-valued current quark mass $m_0$ at zero temperature. In particular, it is shown in the framework of the NJL model under consideration that in the mean-field approximation and under the above restrictions (electric charge neutrality, etc) the chiral imbalance leads to generation of the charged PC phase in quark matter with nonzero baryon density. For this one need to have non-zero chiral isospin chemical potential, i.e. $\mu_{I5}\ne 0$. The non-zero value of chiral chemical potential $\mu_{5}\ne 0$ greatly facilitate this effect and leads to rather large parameter space of dense quark matter occupied by pion condensation phase. So one can say that pion condensation is inevitable phenomenon in dense quark matter with chiral imbalance of both forms,  i.e. at  $\mu_{I5}\ne 0$ and $\mu_{5}\ne 0$. And the regions of pion condensation phase in this case are rather large and located in such a way that it is hard to imagine how other physical constraints on isospin imbalance can avoid this phase and this conclusion can be probably generalized to the conditions realized in heavy ion collisions or neutron star mergers. The investigations have been performed both in the chiral limit and at physical value of current quark mass $m_0$ and it was shown that non-zero value of $m_0$ does not influence much the generation of pion condensation in dense quark matter with chiral imbalance.

The paper is organized as follows. First, in Sec. II a (3+1)-dimensional NJL model
with two quark flavors ($u$ and $d$ quarks) that includes four kinds of chemical potentials, $\mu_B,\mu_I,\mu_{I5}$ is introduced. Then, the condition of electric charge neutrality and $\beta$-equilibrium are discussed and corresponding thermodynamic potential of the model is obtained in the mean-field approximation. The phase structure of the system is discussed  in Sec. III. Sec. III A contains study of the phase structure of the considered model without chiral imbalance. In Sec. \ref{IIIB1} two cases, with $\mu_{I5}\ne 0$, $\mu_{5}=0$ and $\mu_{I5}=0$, $\mu_{5}\ne 0$, are considered and different phase portraits of the model are depicted in the chiral limit. Then in Sec \ref{IIIB2} this discussion is generalized to the case when $\mu_{I5}\ne 0$ and $\mu_{5}\ne 0$, also in the chiral limit. Finally, in Sec. III C the influence of $m_0=5.5$ MeV on the phase structure of the system under consideration is discussed. In particular, it is shown here that at rather high values of $\mu_{I5}$ and  $\mu_{5}$, larger than the mass of $\pi$ meson, the nonzero $m_0$ does not influence significantly on the results obtained in the chiral limit. Hence, in dense quark matter under electric charge neutrality and $\beta$-equilibrium constraints the charged PC phase can be realized, if there is chiral imbalance in the system (both at $m_0=0$ and $m_0\ne 0$), in addition.
Sec. IV contains summary and conclusions. 

\section{The model}

As a starting point of our study, let us consider an auxiliary (3+1)-dimensional system, which is composed of $u$ and $d$ quarks and electrons and described by the following Lagrangian:
\begin{eqnarray}
&&  L=\bar q\big [\gamma^\nu\mathrm{i}\partial_\nu-m_0 \big ]q+ G\Big [(\bar qq)^2+(\bar q\mathrm{i}\gamma^5\vec\tau q)^2 \Big ]
+\bar\psi\big (\gamma^\nu\mathrm{i}\partial_\nu-m_{e} \big )\psi. \label{1}
\end{eqnarray}
In Eq. (1) $q(x)$ is the flavor doublet, $q=(q_u,q_d)^T$, where $q_u$ and $q_d$ are four-component Dirac spinors corresponding to $u$ and $d$ quarks, as well as a color $N_c$-plet, and $\psi(x)$ is also a Dirac spinor but of electrons (the summation in (\ref{1}) over flavor and spinor indices is implied); $\tau_k$ ($k=1,2,3$) are Pauli matrices. For simplicity, we assume that quarks and electrons do not interact with each other and, in addition, we will assume in the following that the mass $m_e$ of electrons is equal to zero. Since the Lagrangian (1) is invariant  with respect to the abelian $U_B(1)$, $U_{I_3}(1)$ and $U_{Q}(1)$ groups,
\begin{eqnarray}
U_B(1):~q\to\exp (\mathrm{i}\alpha/3) q;~
U_{I_3}(1):~q\to\exp (\mathrm{i}\alpha\tau_3/2) q;~
U_{Q}(1):~q\to\exp (\mathrm{i}\widetilde Q
\alpha) q,~~~\psi\to\exp (\mathrm{i}e
\alpha)\psi,
\label{20}
\end{eqnarray}
where $\widetilde Q={\rm diag}(2|e|/3,-|e|/3)$ and $e$ is the value of electric charge of an electron, there are three conserved charges (quantities) in the system (1), total baryon charge, third component of isotopic spin and total electric charge. The density operators $n_B$, $n_I$ and $n_Q$ of these conserved quantities have, respectively, the form
\begin{eqnarray}
&&n_B\equiv\frac 13\bar q\gamma^0q=\frac 13(n_u+n_d),~~ n_I\equiv\frac 12\bar q\gamma^0\tau^3 q=\frac 12(n_u- n_d),\nonumber\\
&& n_{Q}\equiv\bar q\widetilde Q\gamma^0q-|e|\bar\psi\gamma^0\psi=\frac{2|e|}{3} n_u-\frac{|e|}{3} n_d-|e| n_e,\label{30}
\end{eqnarray}
where $ n_u=\bar q_u\gamma^0q_u$, $ n_d=\bar q_d\gamma^0q_d$ and $ n_e=\bar\psi\gamma^0\psi$ are the operators of the particle number densities of $u$ and $d$ quarks and electrons, respectively.  From now on we suppose that $|e|=1$. As a result, it is clear that in the framework of the model (1) it is possible to study some properties of dense matter, composed of $u,d$ quarks and electrons, with fixed values of baryon charge, third component of the isotopic spin and total electric charge. We also assume that it is in the state of thermodynamic equilibrium at zero temperature. It is well known that a consideration of this dense matter is more convenient to perform in terms of chemical potentials, 
which are the quantities, thermodynamically conjugated to corresponding densities $n_B$, $n_I$ and $n_Q$ presented in Eq. (\ref{30}). Therefore, when studying  such a dense medium, one could rely on the Lagrangian of the form
\begin{eqnarray}
  \bar L&=&L+\mu_Bn_B+\mu_In_I+\mu_Qn_Q\nonumber\\
&=&L+\mu_un_u+\mu_dn_d+\mu_en_e,
 \label{40}
\end{eqnarray}
where $L$ is presented in Eq. (1) and $\mu_B$, $\mu_I$ and $\mu_Q$ are baryon-, isospin (isotopic)- and electric charge chemical potentials of the system, respectively. Moreover, in the second line of Eq. (\ref{40}) we introduce the particle number chemical potentials $\mu_u,\mu_d$ and $\mu_e$ of $u$ and $d$ quarks and electrons, respectively, where
\begin{eqnarray}
\mu_u=\frac{\mu_B}{3}+\frac{\mu_I}{2}+\frac{2\mu_Q}{3},~~\mu_d=\frac{\mu_B}{3}-\frac{\mu_I}{2}-
\frac{\mu_Q}{3},~~\mu_e=-\mu_Q.
 \label{04}
\end{eqnarray}
As it was noted above, we will study dense cold matter, described by the Lagrangian (\ref{40}), under two constraints. The first is the reqiurement that matter remains in $\beta$ equilibrium, i.e., all $\beta$ processes such as $d\rightarrow u+e+\bar\nu_{e}$, $u+e\rightarrow d+\nu_{e}$, etc should go with equal rates in both directions. This means that there should hold the relation $\mu_d=\mu_u+\mu_e-\mu_{\nu}$ between chemical potentials of corresponding particles. Since all neutrinos ($\bar\nu_{e}$ and $\nu_{e}$) leave the system, we assume that their chemical potential $\mu_{\nu}$ is zero. So, as a consequence of $\beta$ equilibrium, we have the following relation
\begin{eqnarray}
\mu_u+\mu_e=\mu_d.
 \label{040}
\end{eqnarray}
Taking into account the relations (\ref{04}), we conclude from Eq. (\ref{040}) that $\mu_I$ is equal to zero, i.e., the Lagrangian (\ref{40}) describes dense quark matter in the $\beta$-equilibrium state only at $\mu_I=0$. In addition, we will also impose on the system under consideration the requirement of local electrical neutrality, $\vev{n_Q}=0$, \footnote{Here and below $\vev{A}$ means the ground state expectation value of the operator $A$.} thereby approximating its properties to the properties of such astrophysical objects as compact (neutron) stars. (In spite of the fact that these objects are electrically neutral as a whole, we impose, for simplicity, a stronger demand. Namely, that the electric neutrality is observed at every point of these systems, i.e. locally.) Recall that dense quark matter restricted by $\beta$-equilibrium and electric charge neutrality requirements and described by the Lagrangian (\ref{40}) with $\mu_I=0$ was investigated in Ref. \cite{eklim1} in the chiral limit. In this case, the existence of the charged PC phase is allowed. But if $m_0=5.5$ MeV, then charged PC phase cannot be realized in electrically neutral dense quark matter with $\beta$ equilibrium \cite{abuki}.

However, it is interesting to note that, according to recent studies \cite{andrianov}, chiral asymmetry can arise in dense quark medium inside compact stars. This phenomenon can be described using  two new additional parameters, chiral $\mu_5$- and chiral isospin $\mu_{I5}$ chemical potentials. Moreover, it was shown in our previous papers \cite{kkz,kkz2} that chiral asymmetry, especially in the presence of $\mu_{I5}\ne 0$, is a reliable basis for the appearance of the charged PC effect in dense quark matter, described in the framework of the model (1)-(\ref{40}). But in \cite{kkz,kkz2} we have used $m_0=0$ and did not take into account the possible electrical neutrality of the medium. Naturally, in the present paper, we fill this gap and investigate the phase structure of cold and dense quark matter in the framework of the Lagrangian
\begin{eqnarray}
&&  {\cal L}=\bar L\big |_{\mu_I=0}+
\frac{\mu_{I5}}2 \bar q\tau_3\gamma^0\gamma^5q+\mu_{5}\bar q\gamma^0\gamma^5q,
 \label{50}
\end{eqnarray}
where $\bar L$ is given in Eq. (\ref{40}), under the requirement that in the ground state of the system, i.e. in the state of thermodynamic equilibrium, the density $n_Q$ of electric charge should be zero. \footnote{In order not to overload the paper with additional notations, in the following we denote the ground state expectation value of any operator in the same way as the operator itself, i.e. in our article $n_Q\equiv \vev{n_Q}$, etc.  } Since $\mu_I=0$ in the Lagrangian (\ref{50}), the corresponding system is in the $\beta$-equilibrium state.

Remember that the 2$\times$2 matrix $\widetilde Q$ of electric charge of quarks (see in Eqs (\ref{20})-(\ref{40})) can be presented in the following form $\widetilde Q=I_{3}+\frac{B}{2}$, where $I_{3}=\tau_{3}/2={\rm diag}(1/2,-1/2)$ is the third component of isospin, $ B={\rm diag}(1/3,1/3)$. Using this relation one can obtain for initial Lagrangian (\ref{50}) the following expression
\begin{eqnarray}
&& {\cal L}=\bar q\Big [\gamma^\nu\mathrm{i}\partial_\nu-m_0
+\bar\mu\gamma^0+\nu\tau_3\gamma^0
+\nu_{5} \tau_3\gamma^0\gamma^5+\mu_{5}\gamma^0\gamma^5\Big ]q
+G\Big [(\bar qq)^2+(\bar q\mathrm{i}\gamma^5\vec\tau q)^2 \Big
]+\bar\psi\Big(\gamma^\nu\mathrm{i}\partial_\nu-\mu_{Q}\gamma^0 \Big )\psi,\label{60}
\end{eqnarray}
where
\begin{eqnarray}
\bar\mu\equiv\mu+\frac{\mu_Q}6,~~\mu\equiv\frac{\mu_B}3,~~\nu\equiv\frac{\mu_Q}2,~~\nu_5\equiv\frac{\mu_{I5}}2.
\label{7}
\end{eqnarray}

Phase structure of dense matter described by Lagrangian (\ref{50})-(\ref{60}) is determined by the values of the so-called order parameters and their behavior vs chemical potentials. In turn, the order parameters are the coordinates of the global minimum point of the thermodynamic potential of the system. Moreover, the ground state expectation values $n_B$,  $n_Q$, $n_5$ and $n_{I5}$ can be found by differentiating the thermodynamic potential of the model (\ref{60}) with respect to the corresponding chemical potentials. The goal of the present paper is the investigation of the ground state properties (or phase structure) of the model (\ref{60}) and clarification of the problem of the existence of a charged PC phase in electrically neutral cold and dense quark matter. At the same time, we pay a special attention to the influence of the chiral asymmetry on the fate of this phase.

To find the thermodynamic potential of the system, we use a semibosonized version of the Lagrangian (\ref{60}), which contains composite boson fields $\sigma (x)$ and $\pi_a (x)$ $(a=1,2,3)$:
\begin{eqnarray}
\widetilde {\cal L}\ds &=&\bar qDq 
-\frac{1}{4G}\left (\sigma^2+\pi_a\pi_a\right )
+ \bar\psi\Big (\gamma^\nu\mathrm{i}\partial_\nu -\mu_{Q}\gamma^0 \Big )\psi,\label{2}
\end{eqnarray}
where
\begin{equation}
D\equiv\gamma^\rho\mathrm{i}\partial_\rho
+\bar\mu\gamma^0
+ \nu\tau_3\gamma^0+\nu_{5}\tau_3\gamma^0\gamma^5+\mu_5\gamma^0\gamma^5-m_{0}-\sigma
-\mathrm{i}\gamma^5\pi_a\tau_a.
\label{5}
\end{equation}
Note that $D$ is a nontrivial operator in coordinate-, spinor- and flavor spaces, but it is proportional to a unit operator in the $N_c$-dimensional color space. (Below in our paper, in all numerical calculations we use $N_c=3$.) In Eq. (\ref{2}) and below the summation over repeated indices is implied. From the semibosonized  Lagrangian (\ref{2}) one gets the equations for the boson fields
\begin{eqnarray}
\sigma(x)=-2 G(\bar qq);~~~\pi_a (x)=-2 G(\bar q \mathrm{i}\gamma^5\tau_a q).
\label{200}
\end{eqnarray}
Note that the composite boson field $\pi_3 (x)$ can be identified with the physical $\pi^0(x)$-meson field, whereas the physical $\pi^\pm (x)$-meson fields are the following combinations of the composite fields, $\pi^\pm (x)=(\pi_1 (x)\mp i\pi_2 (x))/\sqrt{2}$.
Obviously, the semibosonized Lagrangian $\widetilde {\cal L}$ is equivalent to the initial Lagrangian (\ref{60}) when using the equations (\ref{200}). Furthermore, it is clear from Eq. (\ref{20}) that the composite boson fields (\ref{200}) are transformed under the abelian isospin $U_{I_3}(1)$ group in the following manner:
\begin{eqnarray}
U_{I_3}(1):~&&\sigma\to\sigma;~~\pi_3\to\pi_3;~~\pi_1\to\cos
(\alpha)\pi_1+\sin (\alpha)\pi_2;~~\pi_2\to\cos (\alpha)\pi_2-\sin (\alpha)\pi_1.
\label{201}
\end{eqnarray}
Starting from the semibosonized Lagrangian (\ref{2}), one can obtain in the mean-field approximation (i.e. in the one-fermion loop approximation) the following path integral expression for the
effective action ${\cal S}_{\rm {eff}}(\sigma,\pi_a)$ of the boson $\sigma (x)$ and $\pi_a (x)$ fields:
$$
\exp(\mathrm{i}{\cal S}_{\rm {eff}}(\sigma,\pi_a))=
  N'\int[d\bar q][dq][d\bar\psi][d\psi]\exp\Bigl(\mathrm{i}\int\widetilde {\cal L}\,d^4 x\Bigr),
$$
where
\begin{equation}
{\cal S}_{\rm {eff}}
(\sigma(x),\pi_a(x))=-\int d^4x\left [\frac{\sigma^2+\pi^2_a}{4G}
\right ]+\tilde {\cal S}_{\rm {eff}}.\label{3}
\end{equation}
The loop contribution to the effective action, i.e. the term $\tilde {\cal S}_{\rm {eff}}$ in (\ref{3}), is given by:
\begin{eqnarray}
\exp(\mathrm{i}\tilde {\cal S}_{\rm {eff}})&=&N'\int [d\bar
q][dq]\exp\Bigl(\mathrm{i}\int\Big\{\bar q Dq\Big\}d^4 x\Bigr)\int [d\bar
\psi][d\psi]\exp\Bigl(\mathrm{i}\int\Big\{\bar\psi (\gamma^\rho\mathrm{i}\partial_\rho -\mu_{Q}\gamma^0 )\psi\Big\}d^4 x\Bigr)\nonumber\\&=&[\Det D]^{N_c}\Det(\gamma^\rho\mathrm{i}\partial_\rho -\mu_{Q}\gamma^0),
 \label{4}
\end{eqnarray}
where $N'$ is a normalization constant. Using the general formula $\Det D=\exp {\rm Tr}\ln D$, one obtains for the effective action (\ref{3}) the following expression
\begin{equation}
{\cal S}_{\rm {eff}}(\sigma(x),\pi_a(x))=-\int
d^4x\left[\frac{\sigma^2(x)+\pi^2_a(x)}{4G}\right]-\mathrm{i}N_c{\rm Tr}_{sfx}\ln D+{\cal S}_{e},
\label{6}
\end{equation}
where the Tr-operation stands for the trace in spinor- ($s$), flavor- ($f$) as well as four-dimensional coordinate- ($x$) spaces, respectively, and
\begin{equation}
{\cal S}_{e}=-i{\rm Tr}_{sx}\ln (\gamma^\rho\mathrm{i}\partial_\rho -\mu_{Q}\gamma^0)
\label{70}
\end{equation}
is the contribution of electrons to the effective action of a whole system.

The ground state expectation values  $\vev{\sigma(x)}$ and $\vev{\pi_a(x)}$ of the composite boson fields are determined by the following saddle point equations,
\begin{eqnarray}
\frac{\delta {\cal S}_{\rm {eff}}}{\delta\sigma (x)}=0,~~~~~
\frac{\delta {\cal S}_{\rm {eff}}}{\delta\pi_a (x)}=0,~~~~~
\label{05}
\end{eqnarray}
where $a=1,2,3$. Just the knowledge of $\vev{\sigma(x)}$ and
$\vev{\pi_a(x)}$ and, especially, of their behaviour vs chemical potentials supplies us with a phase structure of the model. It is clear from (\ref{201}) that if $\vev{\pi_1(x)}\ne 0$ and/or $\vev{\pi_2(x)}\ne 0$, we have in the system a spontaneous breaking of the isospin $U_{I_3}(1)$ symmetry (\ref{20}). Since in this case the ground state expectation values, or condensates, both of the field $\pi^+(x)$ and of the field $\pi^-(x)$ are not zero, this phase is usually called the charged PC phase. In addition, it is easy to see from (\ref{200}) that the nonzero condensates $\vev{\pi_{1,2}(x)}$ (or $\vev{\pi^\pm(x)}$) are not invariant with respect to the electromagnetic $U_Q(1)$ transformations (\ref{20}) of the flavor quark doublet. Hence in the charged PC phase the electromagnetic $U_Q(1)$ invariance of the model (1) is also broken  spontaneously, and superconductivity is an unavoidable property of the charged PC phase. It is also possible to show that at $m_0\ne 0$ the relations $\vev{\pi_3(x)}= 0$ and $\vev{\sigma(x)}\ne 0$ are fulfilled for arbitrary solution of the saddle point equations (\ref{05}) (this fact has been proved, e.g., in Ref. \cite{ekzf}, where  the similar stationary equations with nonzero bare quark mass were investigated (see Eqs. (14)-(15) of \cite{ekzf})). As a result, the phase with $\vev{\sigma(x)}\ne 0$ and $\vev{\pi_a(x)}=0$ ($a=1,2,3$) we will call the phase of normal quark matter, i.e. the phase in which charged PC is absent.

In the present paper we suppose that in the ground state of the system, i.e. in the state of thermodynamic equilibrium, the ground state expectation values $\vev{\sigma(x)}$ and $\vev{\pi_a(x)}$ do not depend on spacetime coordinates, so
\begin{eqnarray}
\vev{\sigma(x)}\equiv M-m_{0},~~~\vev{\pi_a(x)}\equiv \pi_a, \label{8}
\end{eqnarray}
where $M$ and $\pi_a$ ($a=1,2,3$) are already constant quantities. In fact, they are coordinates of the global minimum point (GMP) of the thermodynamic potential (TDP) $\Omega (M,\pi_a)$.
In the mean-field approximation it is defined by the following expression:
\begin{equation}
\int d^4x \Omega (M,\pi_a)=-{\cal S}_{\rm {eff}}\big (\sigma(x),\pi_a (x)\big )\Big|_{\sigma
(x)=M-m_0,\pi_a(x)=\pi_a},\label{08}
\end{equation}
where ${\cal S}_{\rm {eff}}$ is introduced in Eq. (\ref{6}).
In what follows we are going to investigate the dependence of the global minimum point of the function $\Omega (\sigma,\pi_a)$ vs chemical potentials. To simplify the task, let us note that due to a $U_{I_3}(1)$ invariance of the model, the TDP (\ref{08}) depends on $\pi_1$ and $\pi_2$ through the combination $\pi_1^2+\pi_2^2$. In this case, without loss of generality, one can put $\pi_2=0$ in (\ref{08}). Moreover, since at $m_0\ne 0$ we have $\vev{\pi_3(x)}=0$ (see the corresponding remark between Eqs. (\ref{05}) and (\ref{8})), it is also possible to put $\pi_3=0$ in (\ref{08}) and study the TDP as a function of only two variables, $M\equiv\sigma+m_0$ and $\Delta\equiv\pi_1$. In this case the TDP (\ref{08}) reads
\begin{eqnarray}
\Omega (M,\Delta)=\Omega_q (M,\Delta;\bar\mu,\nu,\nu_5,\mu_5)+\Omega_e,\label{07}
\end{eqnarray}
where the contribution of electrons, $\Omega_e$, to the whole TDP has the form
\begin{eqnarray}
\Omega_{e}=\frac{i{\rm
Tr}_{sx}\ln (\gamma^\rho\mathrm{i}\partial_\rho -\mu_{Q}\gamma^0)}{\int d^4x}=-\int_{0}^{\infty}\frac{d^{3}p}{(2\pi)^{3}}(|\mu_{Q}-|\vec p||+|\mu_{Q}+|\vec p||)=-\frac{\mu_{Q}^{4}}{12\pi^2}\label{06}
\end{eqnarray}
(note that in the final expression (\ref{06}) for the $\Omega_{e}$ we have ignored an inessential infinite constant). The quark contribution to the expression (\ref{07}), $\Omega_q (M,\Delta;\cdots)$, looks like
\begin{eqnarray}
\Omega_q (M,\Delta;\bar\mu,\nu,\nu_5,\mu_5)&=&\frac{(M-m_{0})^2+\Delta^2}{4G}+\mathrm{i}N_c\frac{{\rm
Tr}_{sfx}\ln D}{\int d^4x}\nonumber\\&=&
\frac{(M-m_0)^2+\Delta^2}{4G}+\mathrm{i}N_c\int\frac{d^4p}{(2\pi)^4}\ln\Det\overline{D}(p),\label{070}
\end{eqnarray}
where
\begin{equation}
\overline{D}(p)=\not\!p +\bar\mu\gamma^0
+ \nu\tau_3\gamma^0+\nu_{5}\tau_{3}\gamma^0\gamma^5+ \mu_{5}\gamma^0\gamma^5-M-\mathrm{i}\gamma^5\Delta\tau_1
\label{500}
\end{equation}
is the momentum space representation of the Dirac operator $D$ (\ref{5}) under the constraints $\sigma+m_0=M,\pi_1=\Delta,\pi_{2,3}=0$. It should be noted that if the parameters $\bar\mu$, $\nu$ and $\mu_Q$ are constrained by the relations (\ref{7}), then the TDP (\ref{07})-(\ref{500}) describes the state of dense quark matter in $\beta$ equilibrium with electrons.

It is evident that the integral term in Eq. (\ref{070}) is an ultraviolet divergent. So we regularize this expression, introducing a three-dimensional cutoff parameter $\Lambda$, and integrate in Eq. (\ref{070}) over the region, for which $|\vec p|<\Lambda$. In the following we will study the behaviour of the global minimum point of the TDP (\ref{07}) vs chemical potentials for a special set of the model parameters,
\begin{equation}
G=5.01\, GeV^{-2},~~~\Lambda=0.65\, GeV,~~~m_0=5.5\,MeV,~~~N_c=3.
\label{fit}
\end{equation}
In this case at zero chemical potentials one gets for constituent quark mass the value $M=301.58$ MeV. The same parameter set has been used, e.g., in Refs. \cite{Buballa:2003qv,eklim, eklim1}. Note also that the integration in Eq. (\ref{070}) can be carried out analytically only at $\Delta=0$ or $M=0$ but the obtained expressions would be rather involved (see the technique elaborated in Refs \cite{kkz,kkz2}). However, for the evaluation of the TDP (\ref{070}) at $\Delta\ne 0$ and $M\ne 0$ it is  necessary to use numerical calculations.

As a result, we see that in order to find the phase portrait of dense quark matter in $\beta$ equilibrium with electrons (in this case the system is described by the Lagrangian (\ref{50})-(\ref{60})), it is necessary to find the GMP of the whole TDP (\ref{07})-(\ref{500}) when the chemical potential parameters $\bar\mu$, $\nu$ and $\mu_Q$ are constrained by the Eq. (\ref{7}). Then, if at some fixed values of the chemical potentials the GMP has the form  $(M_0\ne 0,\Delta_0\ne 0)$, the charged PC phase is realized in the system. Whereas the GMP of the form $(M_0\ne 0,\Delta_0= 0)$ corresponds to a normal quark matter (NQM) phase, if $M_0\gg m_0$. However, if in the GMP we have $M_0\approx m_0$, the corresponding phase is called approximately symmetrical (ApprSYM). \footnote{Note that in the chiral limit, i.e.  when $m_0=0$ and the Lagrangian (\ref{60}) is invariant under the chiral transformations, in addition to (\ref{20}), the NQM phase passes to the chiral symmetry breaking (CSB) phase, which also corresponds to the GMP of the form  $(M_0\ne 0,\Delta_0= 0)$. Whereas the GMP of the ApprSYM phase is transformed in this limit to the point $(0,0)$ corresponding to the so-called symmetrical (SYM) phase of the system.} It is evident that the coordinates $M_0$ and $\Delta_0$ of the GMP depend on chemical potentials and satisfy to the following stationary equations
\begin{eqnarray}
\frac{\partial\Omega (M,\Delta)}{\partial M}=0,~~~~~
\frac{\partial\Omega (M,\Delta)}{\partial \Delta}=0.
\label{050}
\end{eqnarray}
In contrast to the bare quark mass $m_0$, in the mean-field approximation the quantity $M_0$ is usually called dynamical quark mass. Below, as a set of independent chemical potentials, which characterize the $\beta$-equilibrium of the system under consideration, we will use, as a rule, the quantities $\mu,\nu,\nu_5$ and $\mu_5$ introduced in Eqs. (\ref{50})-(\ref{7}). Then, if the GMP of the TDP (\ref{07}) under the constraints (\ref{7}) is found for each fixed set $(\mu,\nu,\nu_{5},\mu_5)$ of the chemical potentials, one can say that the most general $(\mu,\nu,\nu_{5},\mu_{5})$-phase portrait of the model (\ref{50}) is established in the $\beta$-equilibrium state. It means that we have found the one-to-one correspondence between any point $(\mu,\nu,\nu_{5},\mu_{5})$ of the four-dimensional space of chemical potentials and possible model phases (NQM, charged PC or ApprSYM phases). However, it is clear that this four-dimensional general phase portrait is quite bulky and it is rather hard to imagine it as a whole. So in order to obtain a more deep understanding of the phase diagram as well as to get a greater visibility of it, it is very convenient to consider different low-dimensional cross-sections of this general $(\mu,\nu,\nu_{5},\mu_{5})$-phase portrait, defined by the constraints of the form $\mu= const$ or $\mu_5=const$, etc.

Since one of the purposes of the present paper is to prove the possibility of the charged PC phenomenon in dense quark matter (at least in the framework of the NJL model (1)) under the $\beta$-equilibrium and electric neutrality conditions, the consideration of the physical quantities $\vev{n_{B}}$ and $\vev{n_Q}$, called baryon- and electric charge densities in the ground state, respectively, are now in order. These quantities are very important characteristics of the ground state. So if the coordinates $M_0$ and $\Delta_0$ of the GMP of the TDP (\ref{07}) is known, then
\begin{eqnarray}
\vev{n_B}\equiv n_B=-\frac{\partial\Omega(M_0,\Delta_0)}{\partial\mu_B}=
-\frac 13\frac{\partial\Omega_q
(M_0,\Delta_0;\bar\mu,\nu,\nu_5,\mu_5)}{\partial\bar\mu}, \label{37}
\end{eqnarray}
\begin{eqnarray}
\vev{n_{Q}}\equiv n_Q=-\frac{\partial\Omega (M_0,\Delta_0)}{\partial\mu_Q}=\frac{\mu_Q^3}{3\pi^2}
-\frac 16\frac{\partial\Omega_q
(M_0,\Delta_0;\bar\mu,\nu,\nu_5,\mu_5)}{\partial\bar\mu}-\frac 12\frac{\partial\Omega_q
(M_0,\Delta_0;\bar\mu,\nu,\nu_5,\mu_5)}{\partial\nu},\label{33}
\end{eqnarray}
where $\Omega_q(M,\Delta;\bar\mu,\nu,\nu_5,\mu_5)$ is the quark contribution (\ref{070}) to the whole TDP (\ref{07}), and the quantities $\mu,\nu,\nu_5$ are introduced in Eq. (\ref{7}).

The quantity $\Det\overline{D}(p)$ from Eq. (\ref{070}) is the eighth-order polynomial over $\eta=p_0+\mu$. Its roots are $\eta_i\equiv\eta(M,\Delta;\nu,\nu_5,\mu_5)$, where $i=1,..,8$. The algorithm for finding the roots $\eta_i$ is presented in Ref. \cite{kkz}. There the details of the calculation of the $\Omega_q(M,\Delta;\bar\mu,\nu,\nu_5,\mu_5)$ are also presented. So the final expression for the TDP (\ref{07}) reeds (at $N_c=3$)
\begin{eqnarray}
\Omega (M,\Delta)&=&\frac{(M-m_0)^2+\Delta^2}{4G}-\frac 32\sum_{i=1}^{8}\int\frac{d^3p}{(2\pi)^3}|\mu+\nu/3-\eta_{i}|-\frac{4\nu^{4}}{3\pi^2}. \label{007}
\end{eqnarray}
Note that the TDP $\Omega (M,\Delta)$ (\ref{007}) describes only the $\beta$-equilibrium state of dense quark matter with electrons. But just this expression can also be used in order to obtain the phase structure of electrically neutral quark matter in $\beta$ equilibrium, when $m_0\ne 0$. Indeed, in this case one should (i) find on the basis of the TDP (\ref{007}) the general $(\mu,\nu,\nu_5,\mu_5)$-phase portrait of the model (\ref{50})-(\ref{60}) and then (ii) select all those chemical potentials that obey to the equation $n_{Q}=0$. As a result, in the general $(\mu,\nu,\nu_5,\mu_5)$-phase diagram of the model (\ref{50})-(\ref{60}) we obtain a three-dimensional manifold of physically  acceptable chemical potential values. Finally, we are going to prove that among all these acceptable chemical potentials one can find those that corresponds to the charged PC phase with $n_B\ne 0$.

But in the chiral limit the task is simplified. Indeed, as it was discussed in Ref. \cite{kkz}, in this case it is enough to study the projections $F_1(M)$ and $F_2(\Delta)$ of the TDP (\ref{007}) on the $M$ and $\Delta$ axes, respectively
\begin{eqnarray}
F_{1}(M)&\equiv& \Omega(M,0)
=\frac{M^2}{4G}-\frac{3}{4\pi^2}\sum_{i=1}^{8}\int_{0}^{\Lambda}d|\vec p||\vec p|^2
|\mu+\nu/3-\eta^{M}_{i}|-\frac{4\nu^{4}}{3\pi^2},\label{F1ref}\\
F_{2}(\Delta)&\equiv& \Omega(0,\Delta)
=\frac{\Delta^2}{4G}-\frac{3}{4\pi^2}\sum_{i=1}^{8}\int_{0}^{\Lambda}d|\vec p||\vec p|^2
|\mu+\nu/3-\eta^{\Delta}_{i}|-\frac{4\nu^{4}}{3\pi^2},\label{F2ref}
\end{eqnarray}
 where the roots are
\begin{eqnarray}
\eta^{M}_{1,2,3,4}&=&\nu \pm\sqrt{M^2+(|\vec p|\pm(\mu_5-\nu_{5}))^2},~~~
\eta^{M}_{5,6,7,8}=-\nu \pm\sqrt{M^2+(|\vec p|\pm(\mu_5+\nu_{5}))^2},\nonumber\\
\eta^{\Delta}_{1,2,3,4}&=&\nu_{5} \pm\sqrt{\Delta^2+(|\vec p|\pm(\mu_5-\nu))^2},~~~~
\eta^{\Delta}_{5,6,7,8}=-\nu_{5} \pm\sqrt{\Delta^2+(|\vec p|\pm(\mu_5+\nu))^2}.
\label{13}
\end{eqnarray}
Then the GMP of the TDP (\ref{007}) in the chiral limit can be found by comparing the least values of the functions (\ref{F1ref}) and (\ref{F2ref}).

\section{Phase diagram }
\subsection{Charged pion condensation in dense quark matter with isospin imbalance only}
 Let us discuss the story with charged pion condensation phenomenon in {\cal dense quark matter} without chiral imbalance, i.e. at $\mu_{I5}=0$ and $\mu_5=0$. It started with the prediction of the possibility of existence of the charged PC phase in dense quark matter  with isospin asymmetry but without the requirements of its $\beta$ equilibrium and electrical neutrality. In the chiral limit, the investigation of such a quark matter was performed in Ref. \cite{eklim} in the framework of the NJL model (1)-(\ref{40}) at $\mu_B\ne 0$, $\mu_I\ne 0$, but at $\mu_Q=0$, i.e. in the case when the contribution of electrons are not taken into account in the model (1). It has been shown in this paper that at nonzero baryon density and not very large isospin imbalance one can observe the charged PC phase in the system (see the $(\mu,\nu)$-phase portrait of Fig. 1, where $\mu=\mu_B/3$ and $\nu=\mu_I/2$). Then it was shown in Ref. \cite{eklim1} that this prediction stays valid also under the conditions that are realized in neutron star matter, namely if one includes into consideration the electric neutrality and $\beta$-equilibrium requirements. In this case the investigation can be performed in terms of the NJL model (\ref{40}) as well, but under the $\beta$-equilibrium (\ref{040}) and electric neutrality $\vev{n_Q}=0$ conditions, which implies the presence of electrons in the system. For simplicity, as very often is done in similar situations, the consideration in Ref. \cite{eklim1} has been also performed in the chiral limit.

Soon after that, it was shown in Ref. \cite{abuki} that if one accurately include into the NJL model (\ref{40}) the electric neutrality and $\beta$-equilibrium conditions, combined with the fact that the current quark mass $m_0$ is small but non-zero, the charged PC phase is completely evaporated from the phase diagram of quark matter. (And there is no any indication that this phenomenon, the charged pion condensation, can take place at least in the NJL model approach.) Indeed, in this case the $\beta$-equilibrium condition (\ref{040}) makes the contribution of the isospin  chemical potential $\mu_I$ vanish in the Lagrangian (\ref{40}) (see the discussion in the previous section). As a result, only two chemical potentials, $\mu_B$ and $\mu_Q$, continue to exist in the model. Then, (i) it is possible to draw the $(\mu,\mu_Q)$-phase diagram of such a truncated (by the condition $\mu_I=0$) NJL model (\ref{40}) at the physical point, i.e. at $m_0=5.5$ MeV. (ii) After that, one can depict on the phase diagram the curve along which the density of electric charge $\vev{n_Q}$ is zero. The result of these two procedures, performed in Ref. \cite{abuki}, is presented in Fig. 2 which shows the $(\mu,\mu_Q)$-phase portrait of the system composed of dense quark matter in $\beta$-equilibrium with electrons. The "neutrality line" in the figure corresponds to the points $(\mu,\mu_Q)$, for which the electric charge density $\vev{n_Q}$ is zero. Since this line does not intersect the PC-region of the phase diagram, one should conclude that in the system under consideration the existence of the charged PC phase is prohibited at the physical point.
\begin{figure}
\includegraphics[width=0.43\textwidth]{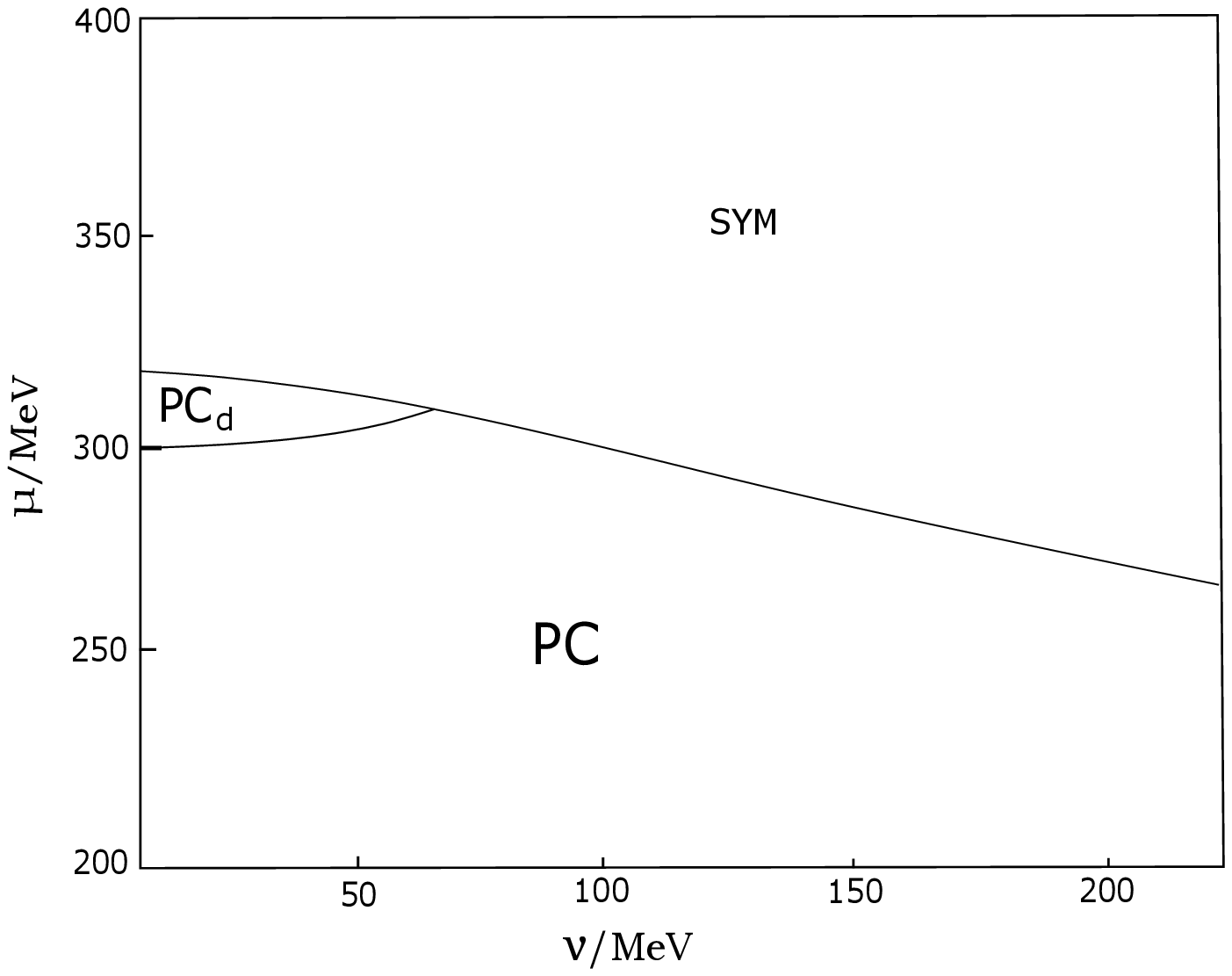}
 \hfill
\includegraphics[width=0.49\textwidth]{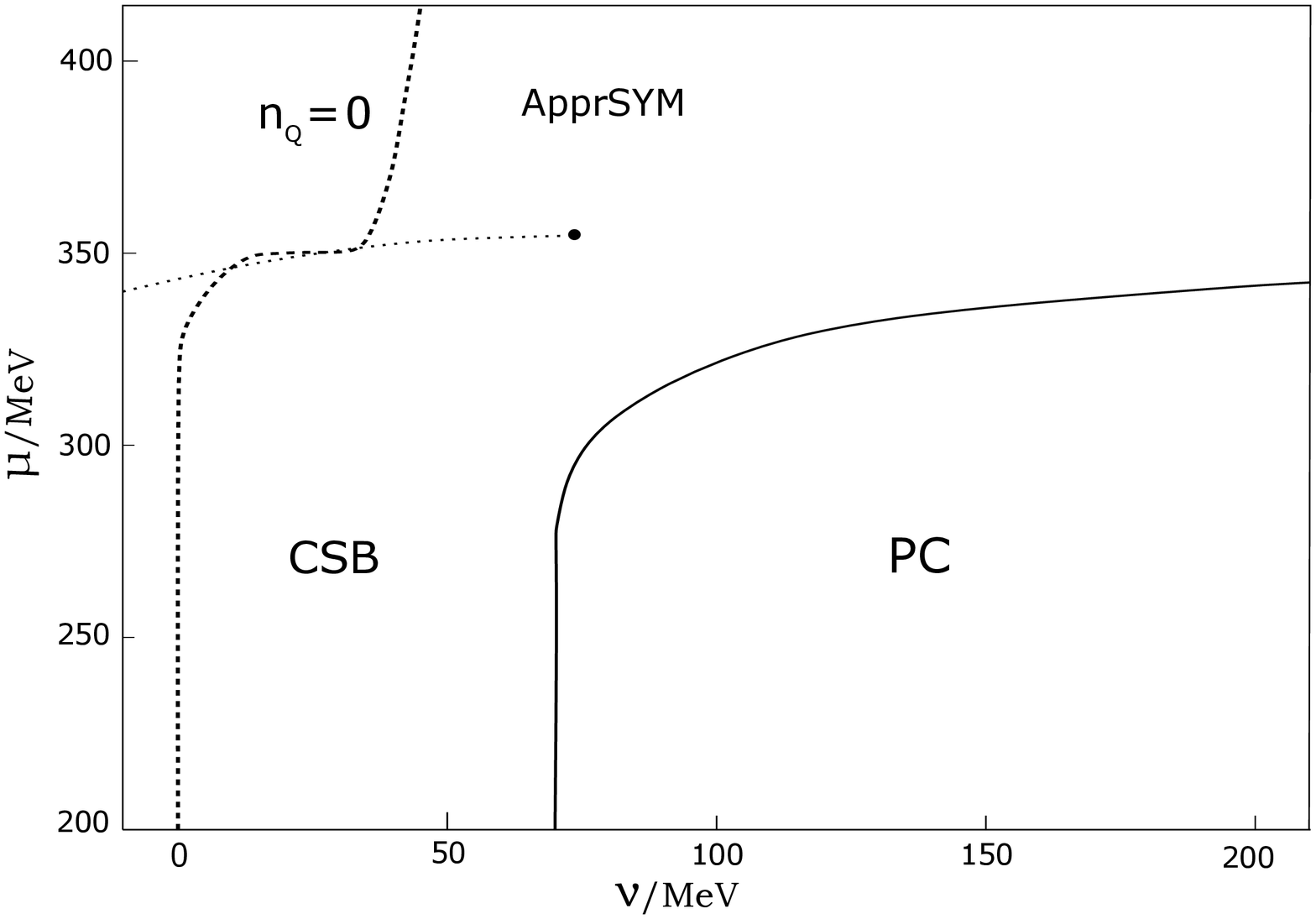}\\
\label{fig1_3}
\parbox[t]{0.45\textwidth}{
 \caption{The $(\mu,\nu)$-phase diagram of the model (\ref{40}) in the chiral limit at $\mu_Q=0$ without charge neutrality and $\beta$-equilibrium conditions (and without chiral asymmetry). Here $\mu=\mu_B/3$, $\nu=\mu_I/2$, PC denotes the charged PC phase with zero baryon density, the PCd denotes the charged PC phase with {\cal nonzero} baryon density. SYM corresponds to symmetrical phase of the model.}
 }\hfill
\parbox[t]{0.45\textwidth}{
\caption{The $(\mu,\mu_{Q})$-phase diagram of dense quark matter which is in the $\beta$-equilibrium with electrons at $\nu_{5}=0$, $\mu_{5}=0$ and $m_0=5.5$ MeV. PC denotes the phase with charged pion condensation. CSB and ApprSYM denote the phases where there is no pion condensation and dynamical quark mass is much greater or around current quark mass value, respectively. Solid and dashed lines are the first order phase transition lines. Dotted line corresponds to  second order phase transitions. Along the dash-dotted "neutrality line" the density of electric charge $\vev{n_Q}$ is zero. } }
\label{fig2}
\end{figure}

Then several years after this consideration there have been found several external conditions that are rather realistic to be realized and that create or enhance the generation of the charged PC phase in the system (see, e.g., in the recent review on this topic \cite{Khunjua:2019nnv}). In particular, it was shown in Ref. \cite{gkkz} that this phase can be realized in dense quark matter, if we allow the  possibility of the existence of spatially inhomogeneous condensates in it. Or take into account that real physical systems have finite sizes \cite{ekkz}. Moreover, it was recently argued that chiral imbalance also can lead to the generation of charged PC phase in dense baryonic (quark) matter, especially if there are two types of the chiral imbalances in it \cite{ekk,kkz,kkz2} (in this case there is even no need for nonzero isospin chemical potential).

Hence,  it would be interesting to check if the charged PC phenomenon can also be realized in electrically neutral dense quark matter in $\beta$-equilibrium with chiral imbalance, i.e. at $\mu_{I5}=2\nu_5\ne 0$ and $\mu_5\ne 0$, or the electric charge neutrality and $\beta$-equilibrium conditions would completely or partially destroy this effect.

\subsection{Accounting for chiral imbalance in the chiral limit, $m_0=0$}\label{IIIB}
To study the problem, it is more convenient to consider from the very beginning the case of zero current quark mass $m_0$. One of the motivations for such an approach is that in this case the expression for the TDP (\ref{007}) is significantly simplified. Moreover, in Ref. \cite{kkz3} it was shown that the charged PC phase appears on the phase diagram of dense quark matter (without electric charge neutrality and $\beta$-equilibrium  requirements) at rather high values of chiral isospin chemical potential $\mu_{I5}\equiv 2\nu_5$, and in this case the influence of the nonzero current quark mass $m_0$ on the location and sizes of the phase is negligible, i.e. in the region of $\nu_5$ that is larger than pion mass, the chiral limit is a good approximation. Below, we will demonstrate that it is also a viable approximation in the case of electrically neutral quark matter in $\beta$ equilibrium.
\begin{figure}
\includegraphics[width=0.49\textwidth]{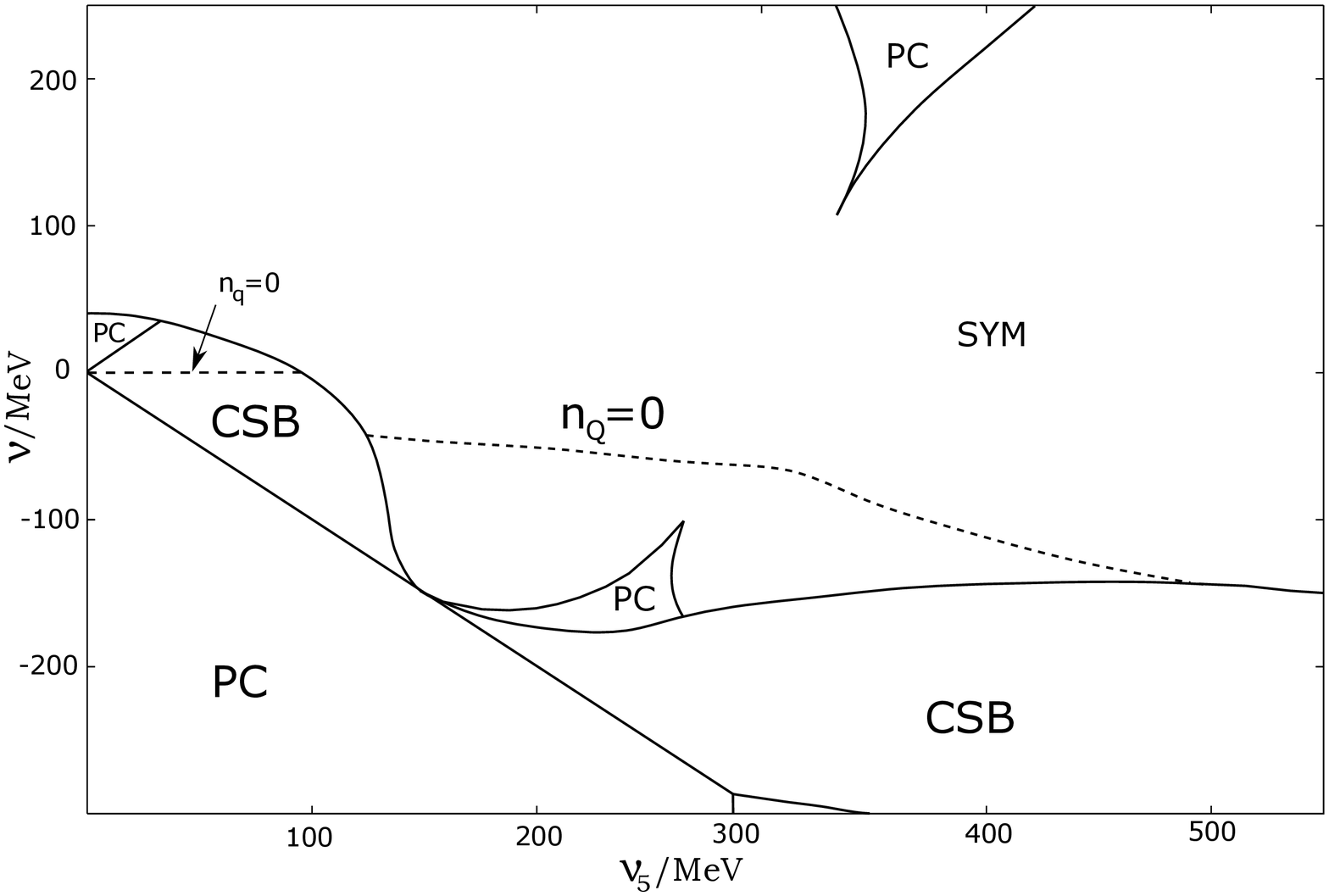}
 \hfill
\includegraphics[width=0.49\textwidth]{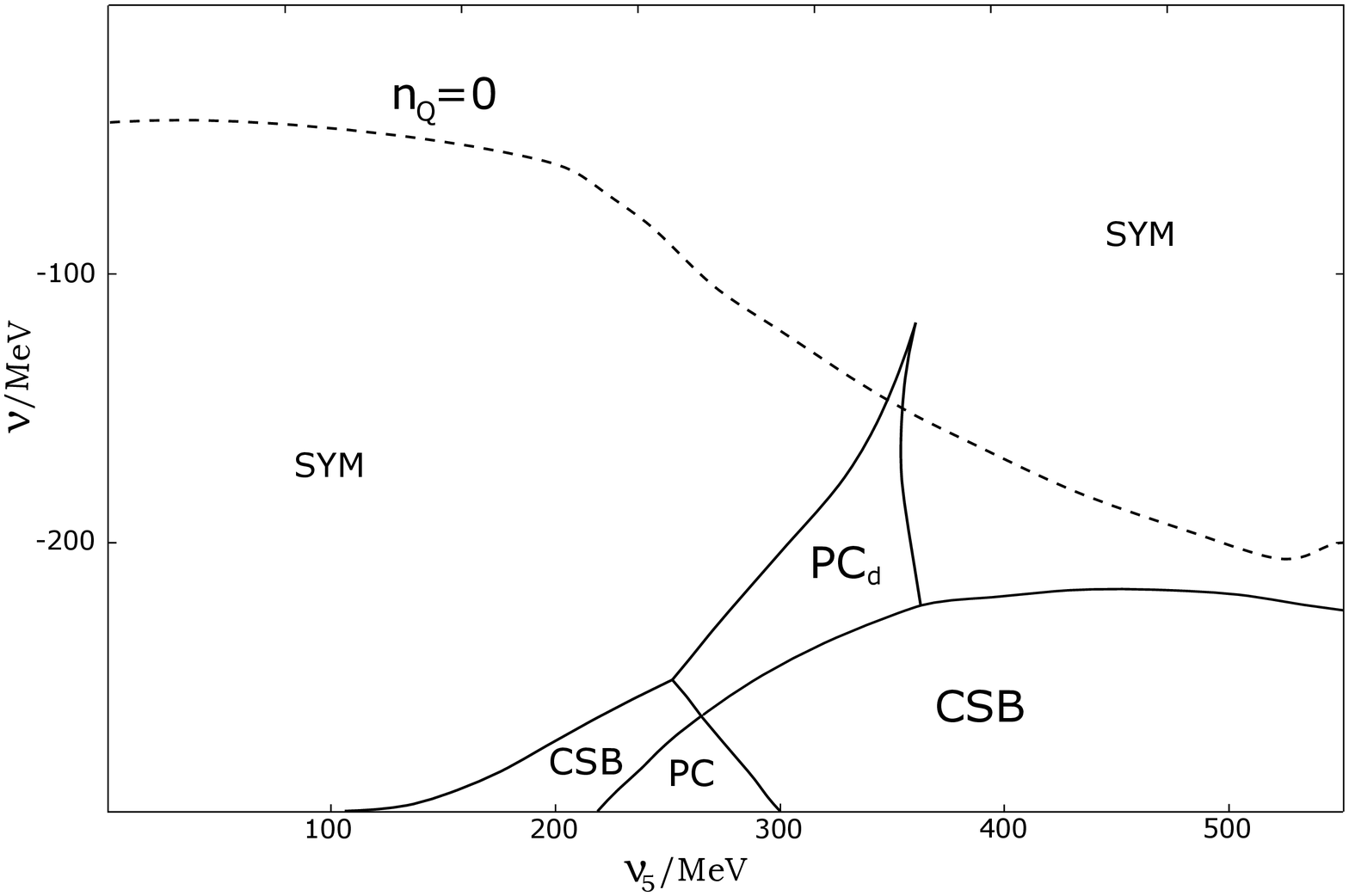}\\
\label{fig1_303}
\parbox[t]{0.45\textwidth}{
 \caption{The $(\nu_{5},\nu)$-phase diagram of $\beta$-equilibrium quark matter (see the model (7)-(8)) at $\mu=300$ MeV and $\mu_{5}=0$ MeV in the chiral limit, $m_0=0$. CSB is the notation for the chiral symmetry breaking phase. Along the dashed line the electric charge density is zero, $n_Q=0$. Other notations are the same as in Fig. 1.}
 }\hfill
\parbox[t]{0.45\textwidth}{
\caption{The $(\nu_{5},\nu)$-phase diagram of $\beta$-equilibrium quark matter (see the model (7)-(8)) at $\mu=400$ MeV and $\mu_{5}=0$ MeV in the chiral limit, $m_0=0$.  Other notations are the same as in Figs. 1 and 3.} }
\label{fig2}
\end{figure}
\begin{figure}
\includegraphics[width=0.49\textwidth]{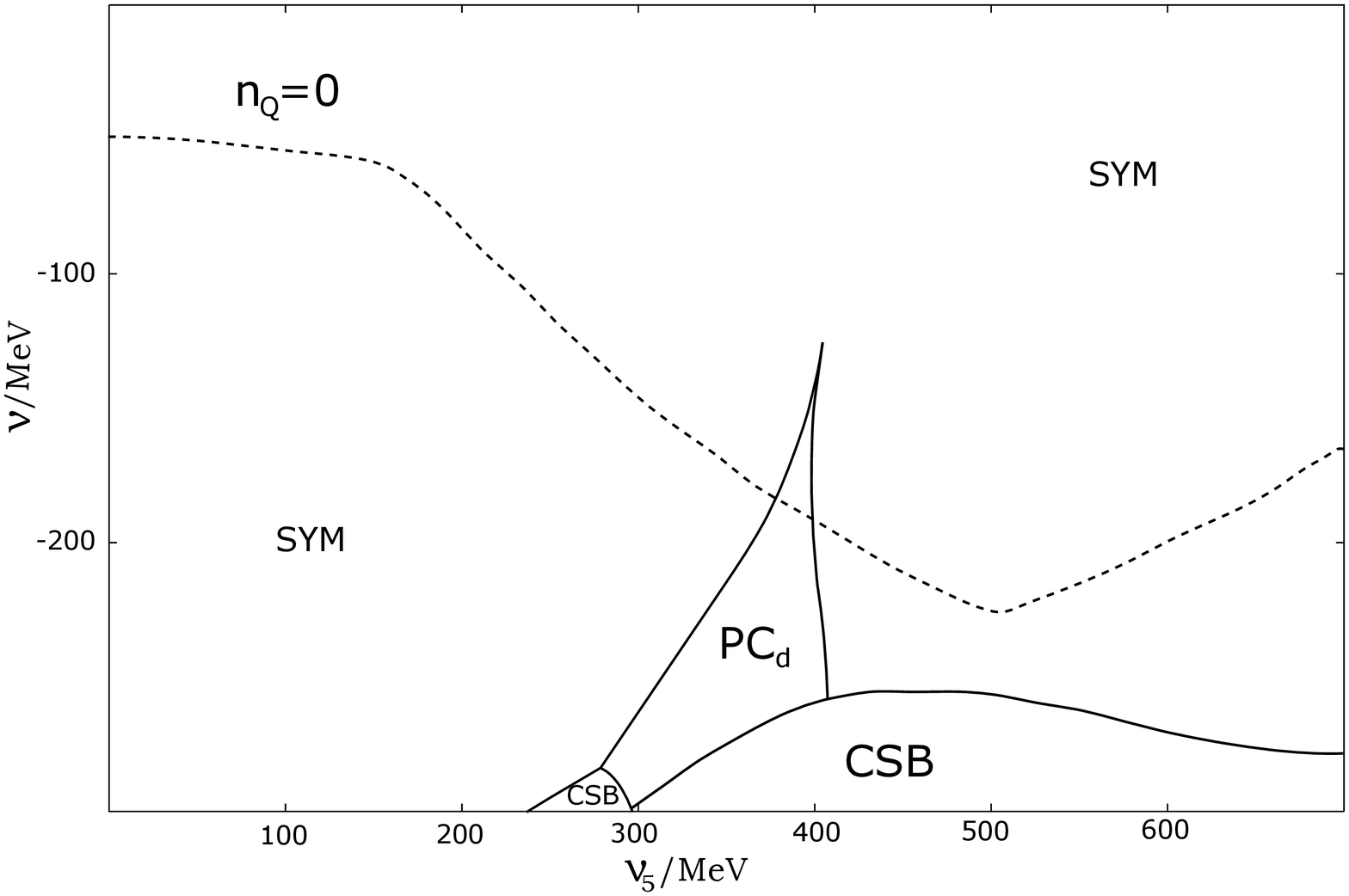}
 \hfill
\includegraphics[width=0.49\textwidth]{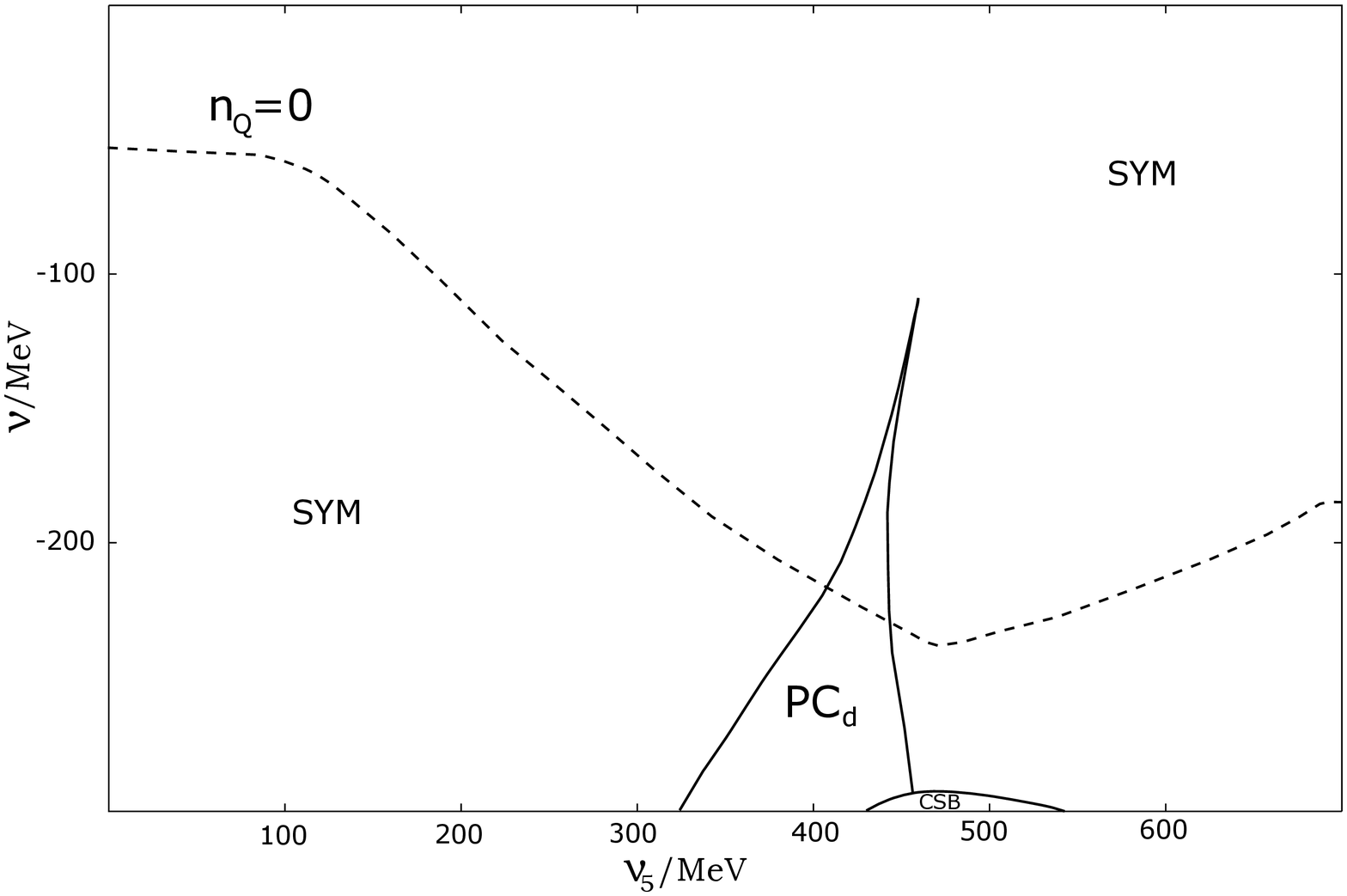}\\
\label{fig7}
\parbox[t]{0.45\textwidth}{
 \caption{The $(\nu_{5},\nu)$-phase diagram of $\beta$-equilibrium quark matter (see the model (7)-(8)) at $\mu=450$ MeV and $\mu_{5}=0$ MeV in the chiral limit, $m_0=0$. All the notations are the same as in Figs. 1 and 3.}
 }\hfill
\parbox[t]{0.45\textwidth}{
\caption{The $(\nu_{5},\nu)$-phase diagram of $\beta$-equilibrium quark matter (see the model (7)-(8)) at $\mu=500$ MeV and $\mu_{5}=0$ MeV in the chiral limit, $m_0=0$. All the notations are the same as in Figs. 1 and 3.} }
\label{fig8}
\end{figure}

\subsubsection{Two particular cases: $\mu_{I5}\ne 0$, $\mu_{5}=0$ and $\mu_{I5}=0$, $\mu_{5}\ne 0$}\label{IIIB1}
Let us first consider dense baryon matter in which chiral imbalance is realized in the particular form, characterized by  $\mu_{I5}\ne 0$, $\mu_{5}=0$.
It means that in this case the $\beta$-equilibrium state of the system with Lagrangian (\ref{50})-(\ref{60}) can be described by three, $\mu$, $\nu$ and $\nu_{5}$, independent chemical potentials.
For simplicity, here the phase properties of this system is represented in terms of different ($\nu_{5},\nu$)-phase diagrams at some fixed values of $\mu$, i.e. we trace its evolution with respect to changing the quark number (or baryon) density of matter.
One can see on the example of the ($\nu_{5},\nu$)-phase diagram at fixed $\mu=0.3$ GeV (see in Fig. 3) that at zero or rather small baryon densities (low values of $\mu$) the line of zero electric charge density $n_{Q}=0$ does not cross the charged PC phase at any values of chiral isospin chemical potential $\nu_{5}$. Hence, at $\mu_5=0$ but at $\nu_5\ne 0$ in the electrically neutral low density quark matter (corresponding to $\mu\leq 300$ MeV) in $\beta$-equilibrium state, only the CSB or symmetrical phase can be realized.
At larger values of $\mu$, as one can see at Figs. 4-6, the electric charge neutrality line $n_{Q}=0$ crosses the PC$_{d}$ phase (it is the notation for the charged PC phase with nonzero baryon density) at values of $\nu_5$ of several hundred MeV. Indeed, using numerical calculations it is possible to show that starting from the value of quark number  chemical potential $\mu=380$ MeV the PC$_{d}$ phase crosses the $n_{Q}=0$ line, but the crossed region of the PC$_{d}$ phase is very tiny, and it is in a small vicinity around $\nu_5\sim 350$ MeV at $\mu=400$ MeV (see in Fig. 4). However, this region increases in size with increasing chemical potential $\mu$. For example, at the value of $\mu=450$ MeV the PC$_{d}$ phase of electrically neutral etc. medium is realized at values of chiral isospin chemical potential $\nu_{5}$ from 370 to 400 MeV (see in Fig. 5). And if $\mu=500$ MeV, then the region of the PC$_{d}$ phase on the line $n_Q=0$ spans in Fig. 6 from values of $\nu_{5}$ of 400 MeV up to 450 MeV (hence, the region is rather large and of order of 50 MeV). These values of chemical potentials are still less than the cut-off and in the scope of validity of the model.

As a result, we see
that at $\nu_5\ne 0$ and $\mu_5=0$ the charged PC phase can be generated in dense and electrically neutral quark matter in $\beta$-equilibrium, if the quark number chemical potential $\mu\geq 380$ MeV
(of course one should stay in the scope of validity of the model (\ref{50})-(\ref{60})).
The larger baryon density (the quark number chemical potential $\mu$),  the larger chiral imbalance (the value of the chiral isospin chemical potential $\nu_5$) we should have in order the PC$_{d}$ phase to be realized in this electrically neutral system in $\beta$ equilibrium at $\mu_5=0$.

Let us now consider the case when the system under consideration is in the chiral imbalance of an opposite form, i.e. at $\nu_{5}=0$ but $\mu_{5}\ne 0$. In this case, but without $\beta$-equilibrium and electric neutrality constraints, dense quark matter was investigated in Ref. \cite{kkz} where it was shown that the charged PC phase can be generated in the system for a not too large region of $\mu_5$ and only at not too large baryon densities. The same is valid for the model (\ref{50})-(\ref{60}) as such, i.e. in each of its $(\mu_5,\nu)$-phase diagram the PC$_{d}$ phase can appear at rather small region of $\mu_5$ and only at not too large values of $\mu$. However, if in addition to $\beta$ equilibrium the electric charge neutrality constraint is imposed on dense quark medium, then in each $(\mu_5,\nu)$-phase portrait of the model (\ref{50})-(\ref{60}) the line of zero electric charge density $n_{Q}=0$ does not cross this not too extensive region of PC$_{d}$ phase. So one can conclude that there is no charged pion condensation in electrically neutral dense quark matter in $\beta$-equilibrium state, if there is chiral imbalance in the form when $\mu_{I5}=0$ but $\mu_5\ne 0$.

\subsubsection{The general case, $\mu_{I5}\ne 0$ and $\mu_{5}\ne 0$}\label{IIIB2}
We saw in Ref. \cite{kkz} that if there is a chiral imbalance in a form of both nonzero chiral isospin $\mu_{I5}$ and chiral $\mu_{5}$ chemical potentials, then the generation of  PC$_{d}$ phase in dense quark matter (without requirement of electric neutrality, etc) is more common and happens in a rather wide region of the chemical potential space and can be realized at any value of isospin chemical potential $\mu_I$. It is evident that this conclusion is also valid in the framework of the model (\ref{50})-(\ref{60}) itself at any values of $\mu_{Q}$, i.e. in the case of only $\beta$ equilibrium of medium.  So one can expect to have a more pronounced generation of charged pion condensation in dense chirally asymmetric quark matter described by the Lagrangian (\ref{50})-(\ref{60}) with $\mu_{I5}\ne 0$ and $\mu_{5}\ne 0$, when electric neutrality condition is taken into account. Now the consideration of this general case is in order.
\begin{figure}
\includegraphics[width=0.49\textwidth]{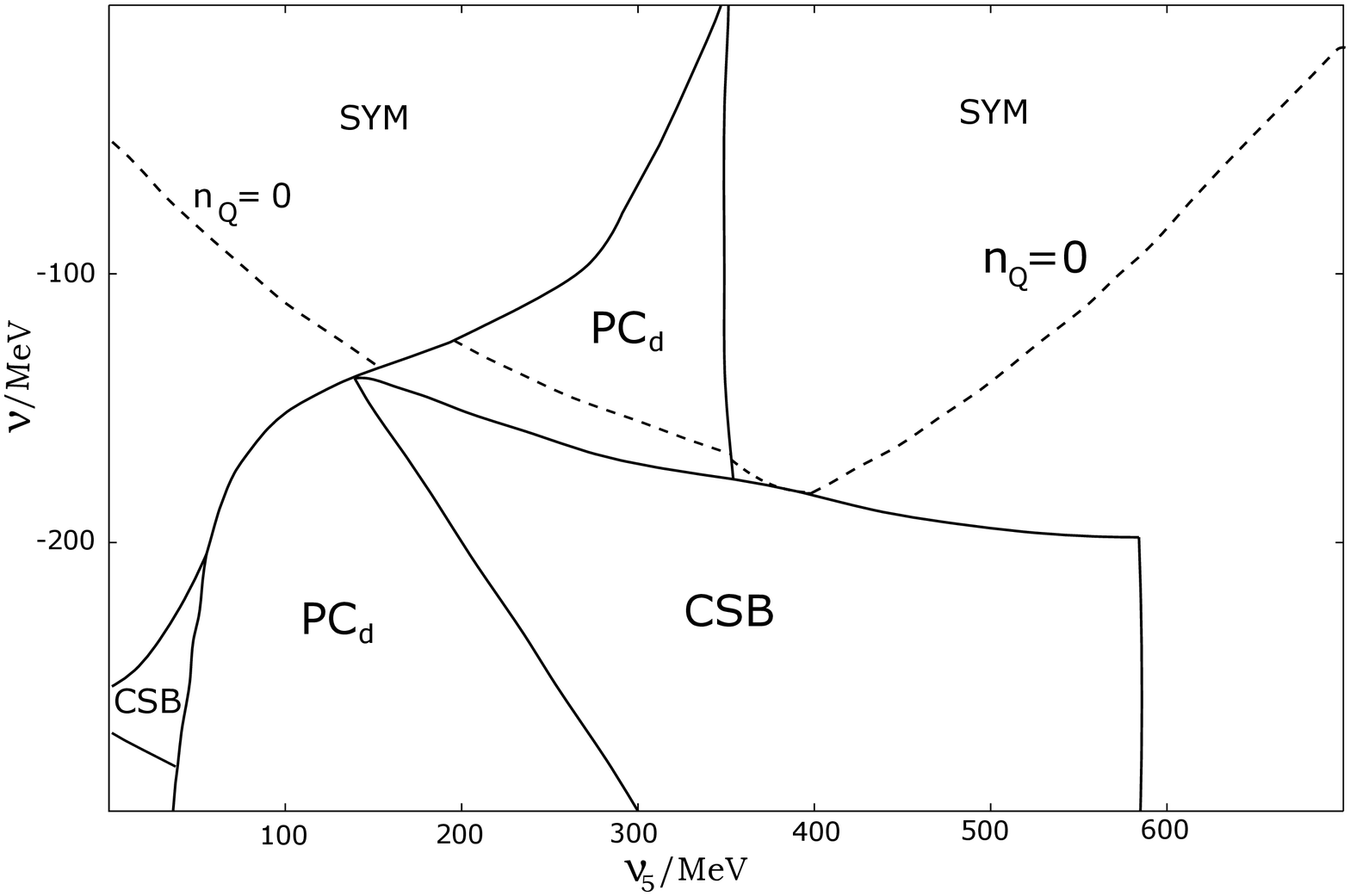}
 \hfill
\includegraphics[width=0.49\textwidth]{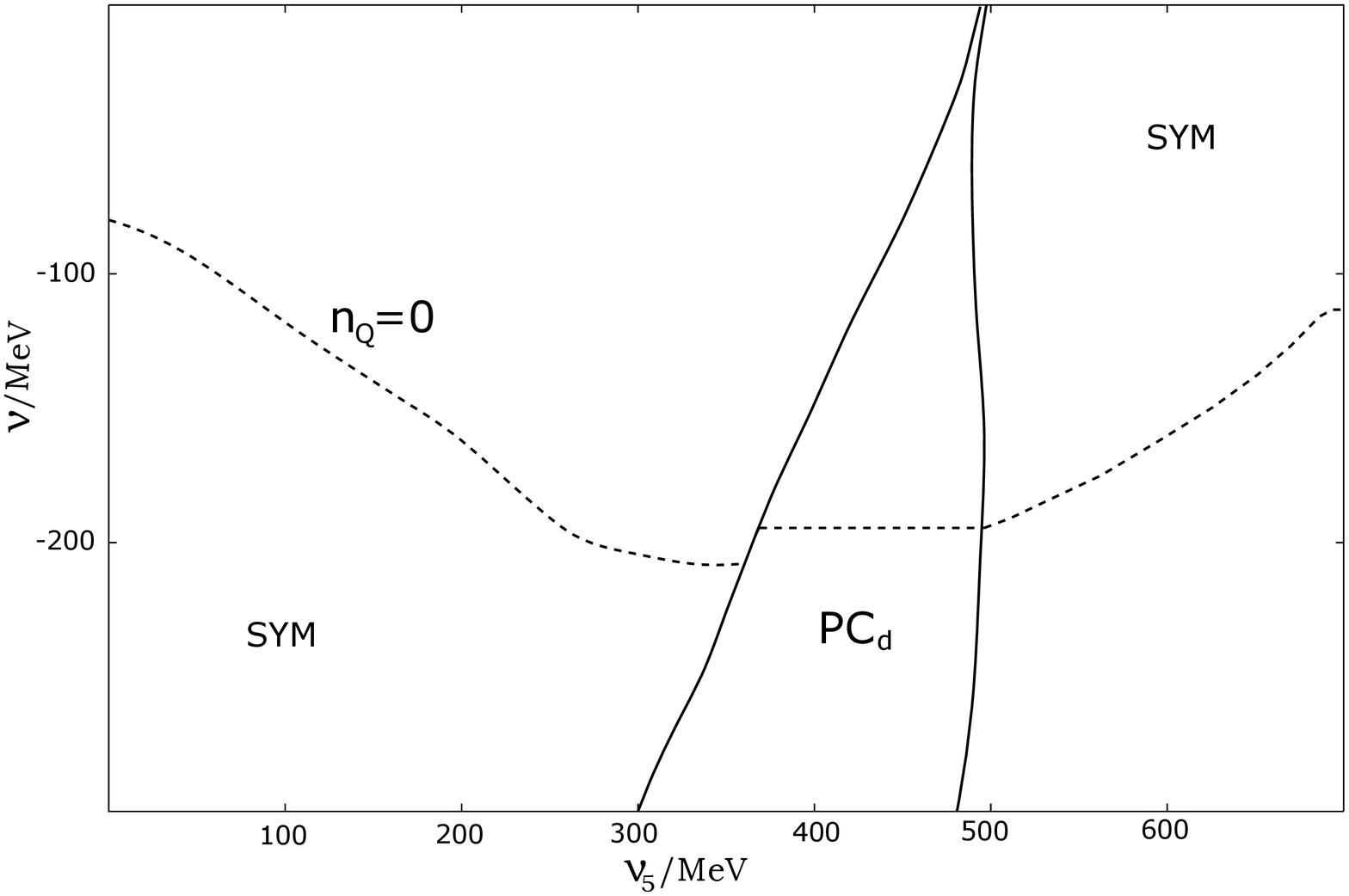}\\
\label{fig14}
\parbox[t]{0.45\textwidth}{
 \caption{The $(\nu_{5},\nu)$-phase diagram of $\beta$-equilibrium  quark matter (see the model (7)-(8)) at $\mu=350$ MeV and $\mu_{5}=150$ MeV in the chiral limit, $m_0=0$. All the notations are the same as in Figs. 1 and 3.}
 }\hfill
\parbox[t]{0.45\textwidth}{
\caption{The $(\nu_{5},\nu)$-phase diagram of $\beta$-equilibrium  quark matter (see the model (7)-(8)) at $\mu=500$ MeV and $\mu_{5}=150$ MeV in the chiral limit, $m_0=0$. All the notations are the same as in Figs. 1 and 3.} }
\label{fig15}
\end{figure}
\begin{figure}
\includegraphics[width=0.49\textwidth]{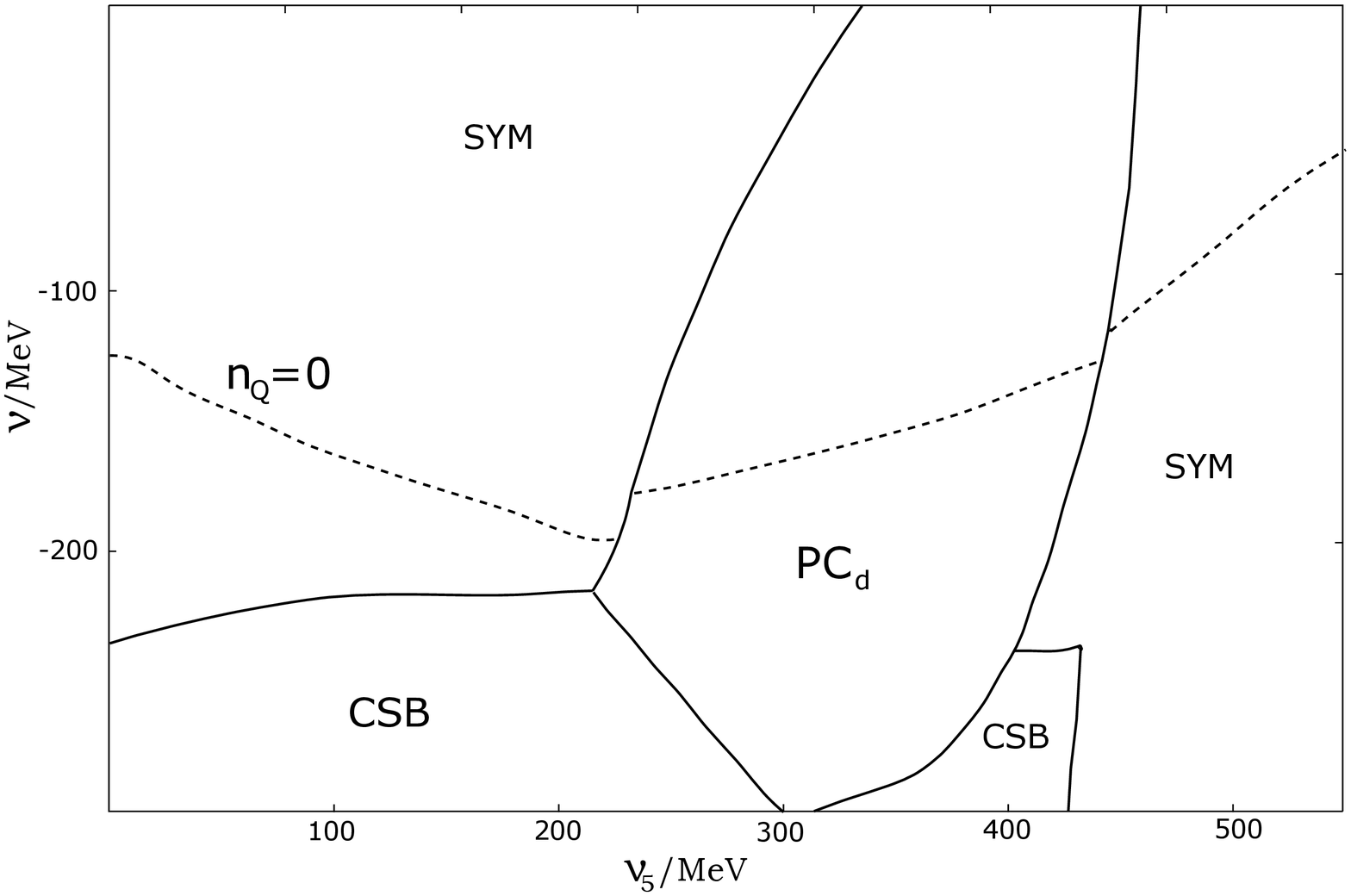}
 \hfill
\includegraphics[width=0.49\textwidth]{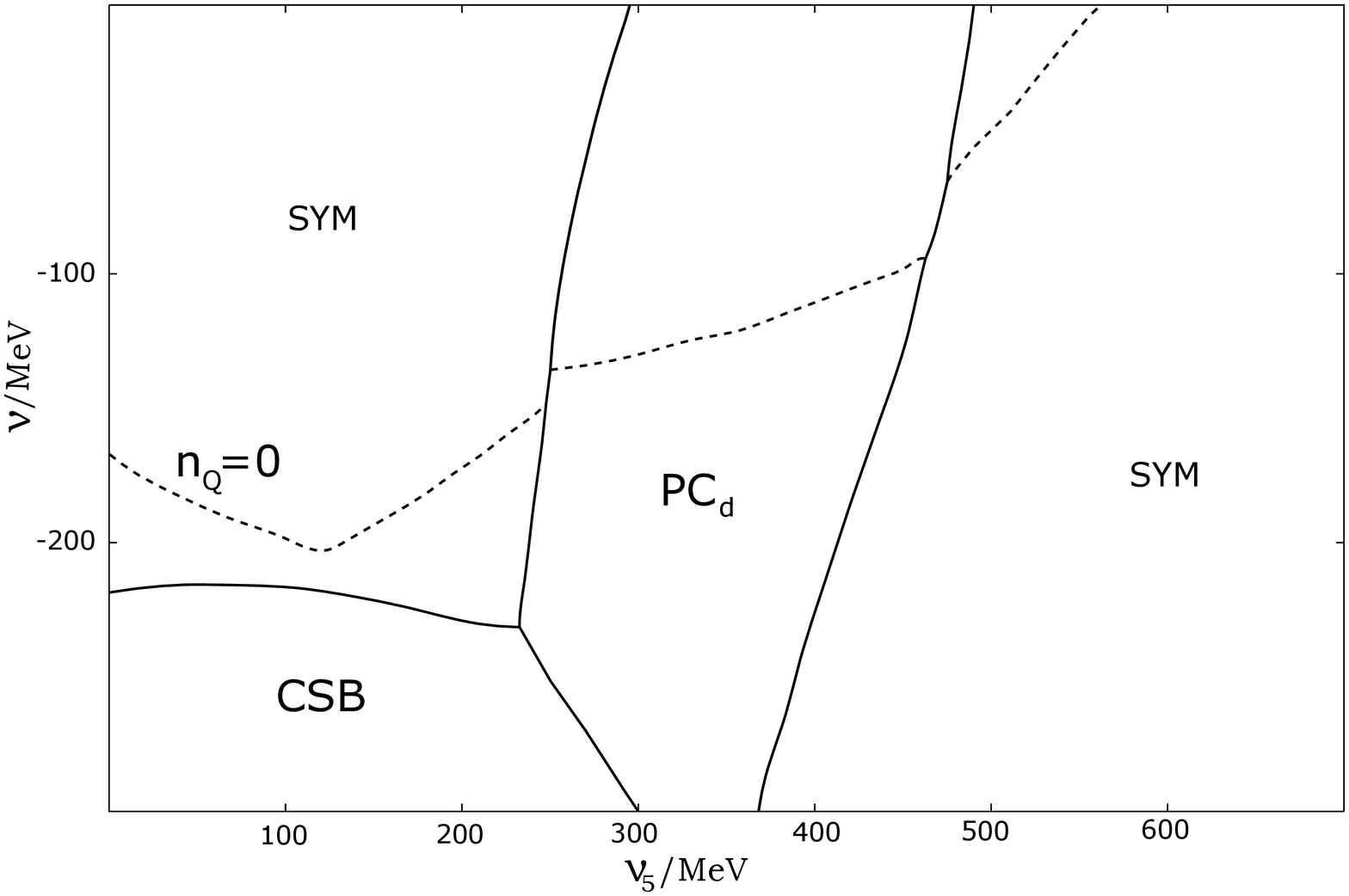}\\
\label{fig7}
\parbox[t]{0.45\textwidth}{
 \caption{The $(\nu_{5},\nu)$-phase diagram of $\beta$-equilibrium  quark matter (see the model (7)-(8)) at $\mu=400$ MeV and $\mu_{5}=300$ MeV in the chiral limit, $m_0=0$. All the notations are the same as in Figs. 1 and 3.} }\hfill
\parbox[t]{0.45\textwidth}{
\caption{The $(\nu_{5},\nu)$-phase diagram of $\beta$-equilibrium  quark matter (see the model (7)-(8)) at $\mu=400$ MeV and $\mu_{5}=400$ MeV in the chiral limit, $m_0=0$. All the notations are the same as in Figs. 1 and 3.} }
\label{fig8}
\end{figure}
The results of this investigation at $m_0=0$ are still formulated in terms of the different $(\nu_5,\nu)$-phase portraits (this time at some fixed values of $\mu$ and $\mu_5$) of the model (\ref{50})-(\ref{60}), in which the electric neutrality is taken into account using the curve $n_Q=0$ (see in Figs. 7-10).

One can see that if $\mu_{5}$ is nonzero, then it is possible to have a rather large region of  PC$_{d}$ phase in dense and electrically neutral quark matter with $\beta$ equilibrium even for sufficiently small values of $\mu$, at which this phase does not realized at $\mu_5=0$. For example, in Fig. 7, corresponding to $\mu=350$ MeV and $\mu_5=150$ MeV, there is a rather spread region (it is of order 150 MeV in length) of the PC$_d$ phase in electrically neutral case, whereas at this value of $\mu$ and at $\mu_5=0$ the PC$_d$ phase is forbidden. Moreover, it is easily seen from these figures that the chemical potential $\mu_5$ promotes a significant increase in the size (from several tens to several hundreds of MeV) of the region of the PC$_d$ phase, compared with the case when $\mu_5=0$. To verify this, it is enough to compare Figs. 6 and 8 for $\mu=500$ MeV with each other, or Figs. 9 and 10 and Fig. 4 for $\mu=400$ MeV, etc.

\subsection{The case of $m_{0}\ne 0$}\label{IIIC}
Let us now prove that in electrically neutral and chirally asymmetric dense quark matter with $\beta$-equilibrium the charged PC phenomenon can be realized not only in the chiral limit, but at the physical point as well, i.e. when current quark mass $m_0=5.5$ MeV. Recall that if chiral asymmetry is absent, then in this case, as it was shown in Ref. \cite{abuki}, the charged PC phenomenon is impossible  at the physical point.  In contrast,  in the chiral limit (and at $\nu_5=\mu_5=0$) this phenomenon takes place in dense quark medium under above mentioned external conditions \cite{eklim1}. So one can ask a question whether the generation of the charged PC$_{d}$ phase by chiral imbalance is smothered by bare quark mass $m_0$ in dense quark matter with charge neutrality and $\beta$-equilibrium. As it was mentioned above (see at the beginning of the section \ref{IIIB}), the inclusion into consideration of a nonzero bare quark mass $m_0=5.5$ MeV does not change significantly both the phase diagram and the behavior of condensates vs. chemical potentials, if chiral isospin chemical potential $\mu_{I5}\equiv 2\nu_5$ of quark matter is rather high in comparison with pion mass $m_\pi$ ($\approx 140$ MeV) \cite{kkz3}. Hence, in this case for a rather wide interval of $m_0$ values both the shape and the sizes of the charged PC$_d$ phase remain approximately the same as this phase has in the chiral limit. But in this paper the possible electric charge neutrality and $\beta$-equilibrium properties of quark medium were not taken into account. Moreover, there only the case $\nu_5\ne 0$, $\mu_5=0$ was studied.

In the paper under consideration we generalize the results of Ref. \cite{kkz3} to the case of chirally asymmetric (with $\nu_{5}\ne 0$ and $\mu_5\ne 0$) dense quark matter under  electric neutrality and $\beta$-equilibrium constraints. But in this case, i.e. at $m_0\ne 0$, the dynamical quark mass $M_0$ (or chiral condensate) is nonzero in all possible phases of the model (\ref{50})-(\ref{60}). Moreover, there is a small chiral condensate $M_0$ even in the charged PC phase, and one cannot use the projections (\ref{F1ref})-(\ref{F2ref}) anymore in order to obtain the $(\nu_5,\nu)$-phase diagrams of the $\beta$-equilibrium quark matter described by the Lagrangians (\ref{50})-(\ref{60}). The calculations become more difficult. Therefore, we abandon the approach used in the previous sections and show numerically (see the examples below) that for each fixed set of chemical potentials $\mu,\nu_5,\mu_5$, which are larger than $m_\pi$, there is a solution $\nu$ of the equation $n_Q=0$ such that the GMP of the TDP  (\ref{007}) corresponds to a charged PC phase. Furthermore, since in this case we have $n_B\ne 0$ in addition, one can conclude that the charged PC phase is easily generated by the chiral imbalance, i.e. at $\nu_{5},\mu_5\gg m_\pi$, in dense quark matter which is in the electrically neutral and $\beta$-equilibrium state even at the physical point ($m_0=5.5$ MeV). Recall once more that without chiral imbalance, i.e. at $\nu_{5}=\mu_5=0$, the existence of the charged PC phase is forbidden in such dense medium (see Ref. \cite{abuki}).

To support this main conclusion of the paper, we present here the relevant numerical investigations for some sets of the chemical potentials $\mu,\nu_5,\mu_5$. For example, in Fig. 11 one can see the behaviors of the GMP coordinates (or condensates) $M_0$ and $\Delta_0$ of the TDP (\ref{007}) vs $\nu$ at $\mu=\nu_5=\mu_5=0.4$ GeV and $m_0=5.5$ MeV. Moreover, the plot of the electric charge density $n_Q$ vs $\nu$ is also depicted in this figure. It is easy to see from the figure that $n_Q=0$ at $\nu\approx -105$ MeV. At this value of $\nu$ the condensates $M_0$ and $\Delta_0$ are nonzero. In addition, in this case we have $n_B\ne 0$. So, for this set of chemical potentials the charged PC phase is realized in electrically neutral dense quark matter with $\beta$ equilibrium. Hence, we see that nonzero $m_0$ cannot suppress the generation of the charged PC phase in this case. For comparison, in Fig. 12 the plots of the same quantities are presented at the same chemical potential values, but in the chiral limit. The Fig. 12 describes the changes of the condensates $M_0$ and $\Delta_0$ and electric charge density $n_Q$ in $\beta$-equilibrium medium, when one moves along a vertical line $\nu_5=400$ MeV on the phase diagram of Fig. 10. Comparing Figs 11 and 12, one can see that there is almost no difference in $\Delta_{0}$ between $m_{0}=0$ and  $m_{0}=5.5$ MeV cases (there is only small difference in $M_{0}$ in the charged PC phase). Moreover, it is clear that the boundary point of the charged PC phase (it is approximately at the point  $\nu\approx -225$ MeV of these figures) in both figures is almost the same, i.e. the size of the charged PC phase does not depend significantly on the bare quark mass $m_0$, thereby generalizing the conclusion of the paper \cite{kkz3} on the case of dense quark matter in $\beta$-equilibrium (at rather large values of $\mu$, $\nu_5$ and $\mu_5$).

In Figs. 13 and 14 the behavior of condensates $M_0$ and $\Delta_0$ as well as electric charge density $n_Q$ vs $\nu$ are depicted for the same values of $\mu$ and $\nu_5$, i.e. for $\mu=400$ MeV, $\nu_5=400$ MeV, but for another value of $\mu_5$, for $\mu_5=300$ MeV, at the physical point and in the chiral limit, respectively.
(In this case the plots of Fig. 14 show the behavior of the condensates $M_0$ and $\Delta_0$ and electric charge density $n_Q$ in $\beta$-equilibrium medium, when one moves along a vertical line $\nu_5=400$ MeV on the phase diagram of Fig. 9.) And it is clear that for this particular choice of $\mu$, $\nu_5$ and $\mu_5$  and at $m_0=5.5$ MeV we can conclude that $n_Q=0$ at $\nu\approx -140$ MeV (see in Fig. 13), where $\Delta_0\ne 0$, i.e. the PC$_d$ phase is realized. Morever, the boundary of the PC$_d$ phase of $\beta$-equilibrium quark matter (in Fig. 13 it is at $\nu\approx - 235$ MeV) is practically the same as in the chiral limit (see in Fig. 14 or in Fig. 9 at $\nu_5=400$ MeV). In a similar way, it is possible to show numerically that, e.g., at  $\mu=400$ MeV and $\nu_5=400$ MeV and for a rather large interval of $\mu_5$ (which certainly contains the region $\mu_5\in (300\div 400)$ MeV) the charged PC$_d$ phase can be realized in electrically neutral and $\beta$-equilibrium quark matter at the physical point.

\begin{figure}
\includegraphics[width=0.49\textwidth]{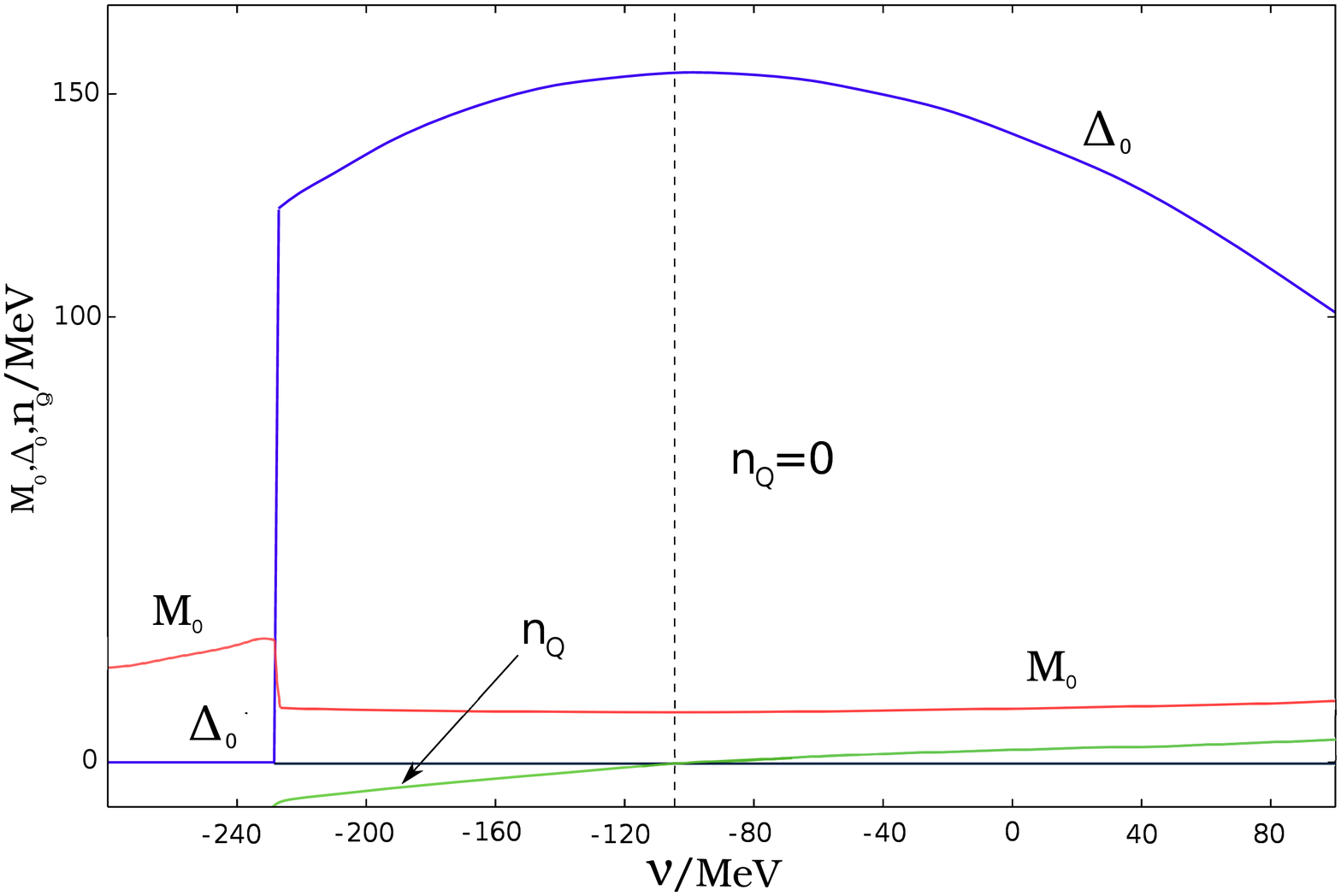}
 \hfill
\includegraphics[width=0.49\textwidth]{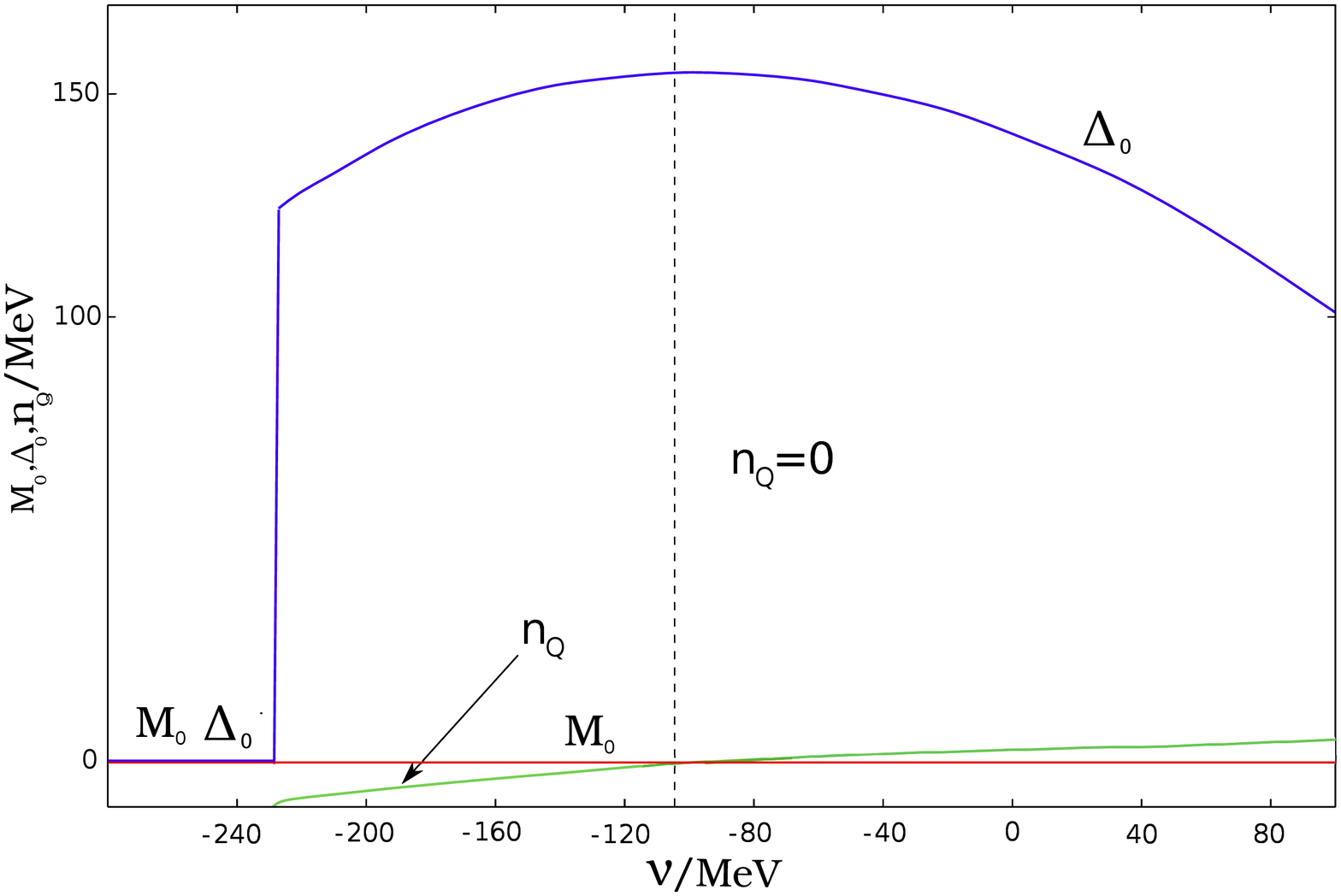}\\
\label{fig14}
\parbox[t]{0.45\textwidth}{
 \caption{  $\Delta_{0},M_{0},n_{Q}$ as a function of $\nu$ at $\mu=400$ MeV and $\mu_{5}=400$ MeV and $\nu_{5}=400$ MeV. The physical point case, current quark mass $m_{0}=5.5$ MeV.  }
 }\hfill
\parbox[t]{0.45\textwidth}{
\caption{ $\Delta_{0},M_{0},n_{Q}$ as a function of $\nu$ at $\mu=400$ MeV and $\mu_{5}=400$ MeV and $\nu_{5}=400$ MeV. The chiral limit case, current quark mass $m_{0}=0$. } }
\label{fig15}
\end{figure}

\begin{figure}
\includegraphics[width=0.49\textwidth]{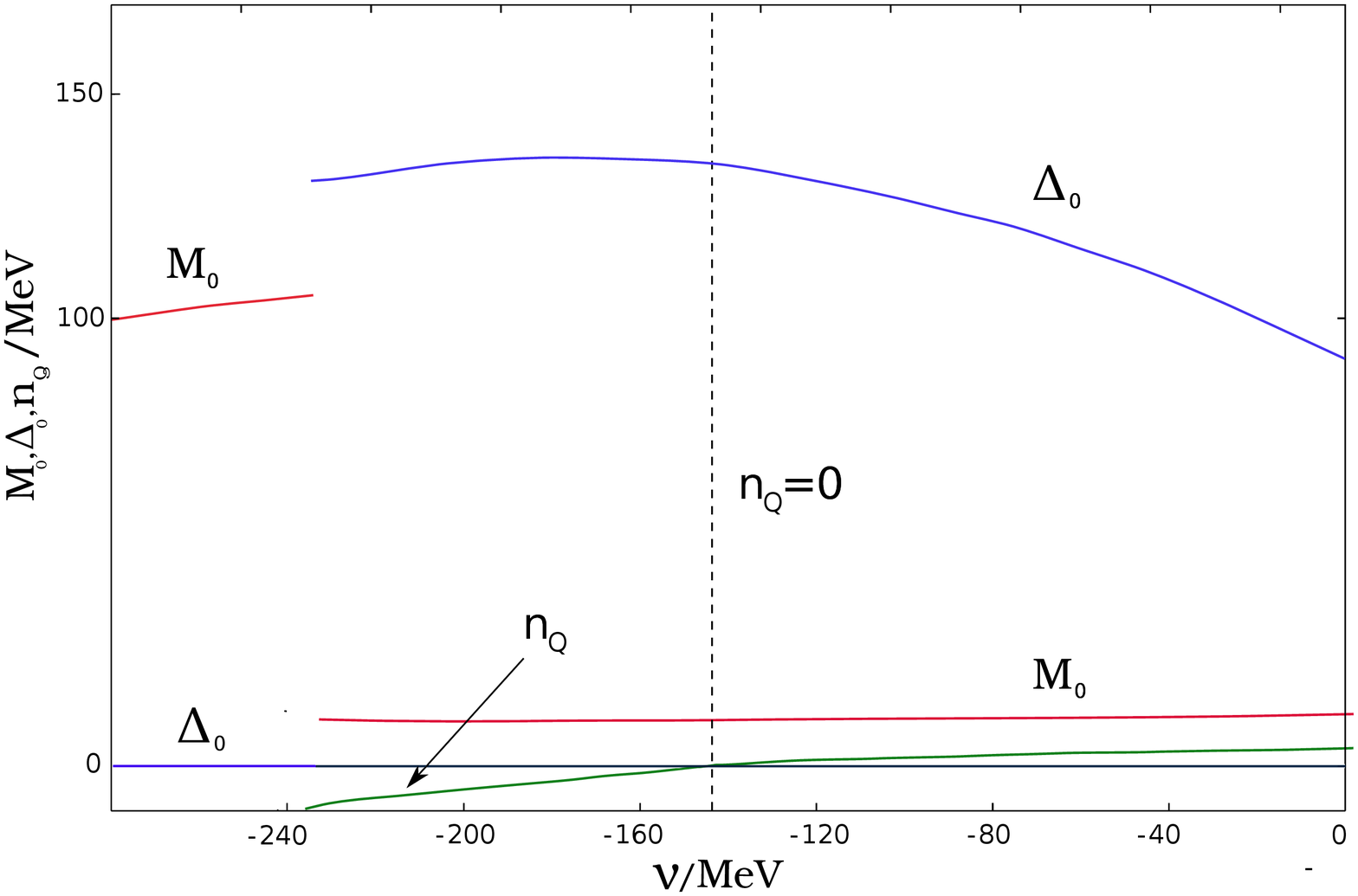}
 \hfill
\includegraphics[width=0.49\textwidth]{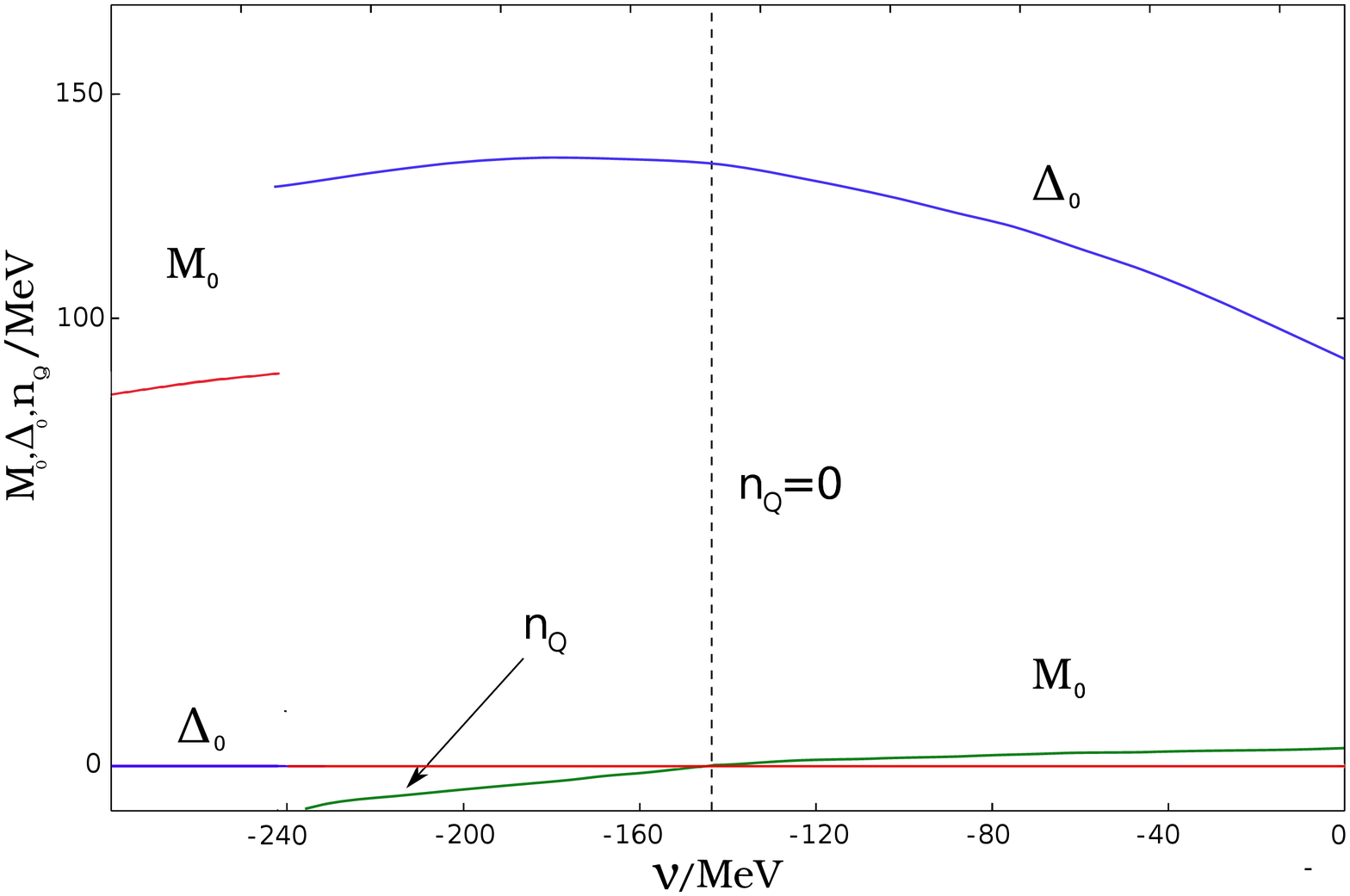}\\
\label{fig12}
\parbox[t]{0.45\textwidth}{
 \caption{  $\Delta_{0},M_{0},n_{Q}$ as a function of $\nu$ at $\mu=400$ MeV and $\mu_{5}=300$ MeV and $\nu_{5}=400$ MeV. The physical point case, current quark mass $m_{0}=5.5$ MeV. }
 }\hfill
\parbox[t]{0.45\textwidth}{
\caption{  $\Delta_{0},M_{0},n_{Q}$ as a function of $\nu$ at $\mu=400$ MeV and $\mu_{5}=300$ MeV and $\nu_{5}=400$ MeV. The chiral limit case, current quark mass $m_{0}=0$. } }
\label{fig13}
\end{figure}
\section{Summary and Conclusions}

In this paper the influence of isotopic, chiral and chiral isotopic imbalances on phase structure of dense quark matter has been investigated in the framework of the mean-field approximation to the (3+1)-dimensional NJL model with two quark flavors under the conditions of electric charge neutrality and $\beta$-equilibrium both with nonzero current quark mass, $m_0\ne 0$, and in the chiral limit, $m_0=0$. Dense matter means that our consideration also has been performed at nonzero baryon $\mu_B$ chemical potential. Isotopic imbalance in the system were accounted for by introducing isospin $\mu_I$ (see Lagrangian (1)) and chiral isospin and chiral imbalances in the system were accounted for by introducing chiral isospin $\mu_{I5}$ and chiral $\mu_{5}$ chemical potentials into the Lagrangian (\ref{50})-(\ref{60}).

Let us remind you the story with prediction of charged pion condensation (PC) in dense baryon matter. The first investigations and predictions that pion condensation could be observed in nuclear matter, such as neutron stars, appeared in the seventies \cite{Migdal}. But soon it was understood that in the framework of effective models for nuclear matter, in which pion is a point particle, the S-wave PC (in this case the PC is a spatially homogeneous) is highly unlikely to be realized. And a possibility for charged P-wave pion condensation has been argued \cite{Tatsumi}. However, in the core of a neutron star the baryon density is several times higher than the  ordinary nuclear density. At such densities, the hadrons can be so close to each other that the boundaries between them are not distinguishable, and in fact we will deal with quark matter (with nonzero baryon density). In this case, to study the properties of a dense baryonic medium, it is more appropriate to use the QCD-like effective models such as the NJL model, etc. But in the NJL models, pions (and other hadrons) are not point particles. They are rather Nambu-Goldstone bosons (in the chiral limit) associated with spontaneous CSB. If there is an isospin imbalance in the system, then, as shown in Refs. \cite{Son,Kogut,Loewe,Brandt, koguthe,eklim,ak}, the formation of S-wave Cooper $\bar u\gamma^5 d$ pairs (with quantum numbers of charged pions) is possible, the condensation of which leads to a spontaneous breaking of the isotopic $U_{I_3}(1)$ invariance (see in Eq. (\ref{20})) and the appearance of the charged (S-wave) PC phenomenon in dense quark matter even under electric charge neutrality and $\beta$-equilibrium conditions (in the chiral limit, $m_0=0$, but without chiral imbalance) \cite{eklim1}. However, in Ref. \cite{abuki} it was shown that in this case, i.e. at $\mu_5=\mu_{I5}=0$, and at the physical point, $m_0=5.5$ MeV, the charged PC phase cannot be realized in electrically neutral dense quark matter.

Then it has been found several conditions, including chiral imbalance at nonzero $m_0$, that can promote this phenomenon in dense baryonic matter \cite{Khunjua:2019nnv,ekkz,gkkz,ekk,kkz,kkz2,kkz3}.
So now it is interesting if the generation of the charged PC can survive rather strict requirement of electric neutrality and $\beta$-equilibrium. Here in this paper the fate of the charged PC phase of dense quark matter is investigated under influence of these external constraints.
Let us summarize the central results of our paper.

1) It is established that charged pion condensation phenomenon is induced by chiral imbalance in dense (i.e. at $\mu_B\ne 0$ when baryon density in nonzero) and electrically neutral medium with $\beta$-equilibrium. This matter exists,  for example, inside neutron stars and generation of charged pion condensation in its cores can lead to a number of potentially interesting physical implications.

2) The discussed generation of charged PC can be promoted even by only chiral isospin chemical potential $\mu_{I5}\equiv 2\nu_5$ at zero chiral imbalance, $\mu_{5}=0$. In this case the generation of this phenomenon takes place in the region of the chemical potential $\nu_5$ which may be 50$\div$70 MeV in size (see in the section \ref{IIIB1}), but perhaps  it is not still so wide and could be easily spoiled by extreme conditions. For example, it is not guaranteed to survive in the environment appearing in heavy ion collisions or neutron star mergers.

3) However, if there are both forms of chiral imbalance in the system (chiral $\mu_{5}$ and chiral isospin $\nu_{5}$ chemical potentials are nonzero), then the generation of charged PC in dense baryon matter, limited by electric charge neutrality, etc, conditions, is a quite inevitable 
phenomenon (see in the section \ref{IIIB2}). It can be realized in a larger region of the parameter space. For example, even for a rather small values of chiral chemical potential $\mu_5\in 50\div 150$ MeV the charged PC phase became much more common in the electrically neutral quark (baryon) matter (it could be in a range of $\nu_{5}$ as large as 150$\div$200 MeV). 

4) It is also can be observed that nonzero values of $\mu_{5}$ rather drastically change the position of the dense charged PC phase, such that it is quite hard to be evaded by any constraints (such as the electric charge neutrality). And this can be probably generalized to the conditions on isospin imbalance from heavy ion collisions or neutron star mergers. Hence, one can probably say that charged PC is an unavoidable phenomenon in dense quark (baryon) matter with two types, $\mu_5\ne 0$ and $\nu_5\ne 0$, of chiral imbalances.

5) Although in dense and electrically neutral quark matter, which is in a $\beta$-equilibrium state but without chiral imbalance (i.e. at $\mu_5= 0$ and $\nu_5= 0$), the physical bare quark mass, $m_0=5.5$ MeV, destroys the charged PC phase observed in the chiral limit, $m_0=0$, \cite{abuki},  the 
inclusion of the physical quark mass into consideration, when chiral imbalance is nonzero, does not spoil the generation of charged PC which is observed in the chiral limit, and even does not almost affect it at all (see the results of the section \ref{IIIC}). Because it happens as a rule at the values of chiral chemical potentials larger than the half of the pion mass. Hence, for these values of $\mu_5$ and $\nu_5$ the consideration in the chiral limit is justified.

It leads to interesting possible applications for physics of neutron stars, since there are several mechanisms of generation of chiral isospin and chiral imbalances in dense matter especially under influence of external magnetic field (see, e.g., the discussion in \cite{kkz2,kkz3}). In a view of latest and forthcoming NICER results and the first observed and possibly new events of neutron star mergers, it is rather interesting to explore how possible chiral imbalance in neutron star can influence and change mass-radius relation, etc.
Here we consider only physical condition pertinent to neutron stars physics, but it can be also an indirect indication that other various physical conditions does not destroy the charged PC phenomenon, and there can be possibility that it happens also in moderate energy  heavy ion collision experiments such as NICA, RHIC and FAIR.

\section{ACKNOWLEDGMENTS}

R.N.Z. is grateful for support of Russian Science Foundation under the grant  19-72-00077. The work was also supported by the Foundation for the Advancement of Theoretical Physics and Mathematics BASIS.


\end{document}